\newcommand{\ce}{\mathop{\mathrm{ce}} \nolimits}
\newcommand{\se}{\mathop{\mathrm{se}} \nolimits}
\newcommand{\Tr}{\mathop{\mathrm{Tr}} \nolimits}

\documentclass[pra,superscriptaddress,eqsecnum,twocolumn,showpacs]{revtex4}
\usepackage{amsmath,amssymb,bm,graphicx,times}

\begin{document}

\title{Experimental test of uncertainty relations for quantum
  mechanics on a circle}

\author{J. \v{R}eh\'{a}\v{c}ek}
\affiliation{Department of Optics,
Palacky University, 17. listopadu
50, 772 00 Olomouc, Czech Republic}

\author{Z. Bouchal}
\affiliation{Department of Optics,
Palacky University, 17. listopadu
50, 772 00 Olomouc, Czech Republic}

\author{R. \v{C}elechovsk\'{y}}
\affiliation{Department of Optics,
Palacky University, 17. listopadu
50, 772 00 Olomouc, Czech Republic}

\author{Z. Hradil}
\affiliation{Department of Optics, Palacky University, 17.
listopadu 50, 772 00 Olomouc, Czech Republic}

\author{L. L. S\'{a}nchez-Soto}
\affiliation{Departamento de \'Optica,
Facultad de F\'{\i}sica,
Universidad Complutense, 28040~Madrid, Spain}

\date{\today}

\begin{abstract}
  We rederive uncertainty relations for the angular position and
  momentum of a particle on a circle by employing the exponential 
  of the angle instead of the angle itself, which leads to circular
  variance as a natural measure of resolution. Intelligent states
  minimizing the uncertainty product under the constraint of a given
  uncertainty in angle or in angular momentum turn out to be given by
  Mathieu wave functions. We also discuss a number of physically 
  feasible approximations to these optimal states. The theory is 
  applied to the orbital angular momentum of a beam of photons and 
  verified in an experiment that employs computer-controlled 
  spatial light modulators both at the state preparation and 
  analyzing stages.
\end{abstract}


\maketitle

\section{Introduction}

Despite the outstanding role that angular variables play in physics,
their proper definition in quantum mechanics is beset by 
difficulties and requires more care than perhaps might be 
expected~\cite{Perinova:1998,Luis:2000,Kastrup:2006}.  Consider, 
for instance, the simple example of a particle moving on a circle 
of unit radius: the problems essentially arise then from the 
periodicity, which prevents the existence of a well-behaved angle 
operator, but not of its complex exponential, we shall denote by 
$\hat{E}$.

In quantum optics, this topic is by no means purely academic: it 
turns out to be crucial for a proper understanding of, e.g., the 
orbital angular momentum (OAM) of light~\cite{Allen:2003}.  Indeed, 
as put forward by Allen and coworkers~\cite{Allen:1992}, the 
Laguerre-Gauss modes, typical of cylindrical symmetry, carry 
a  well-defined OAM per photon.  Since it is surprisingly 
simple  to  generate, control, filter, and detect OAM states 
of light  experimentally, researchers have begun to appreciate 
their  practical potential for classical~\cite{Bouchal:2004,
Gibson:2004,Celechovsky:2007} and  quantum information 
applications~\cite{Mair:2001,Vaziri:2002,Molina:2004,Langford:2004,
Oemrawsingh:2005,Marrucci:2006,Molina:2007}.

Intimately linked to the issue of a proper angle description it is the
question of the associated uncertainty relations. Surprisingly enough,
some subtle aspects of these relations still remains under discussion.
From previous work in this topic~\cite{Nieto:1967,Zak:1969,Whelan:1980,
Loss:1992,Ohnuki:1993,Szabo:1994,Kostelecky:1996,Kowalski:2002,Bang:2006} 
it seems clear that, if one insists in holding to an angle operator, 
special care must be  taken when using the standard variance, since 
this is a nonperiodic measure of spread that makes the angular 
uncertainty  depend on the $2\pi$ window chosen.  Moreover, the 
associated commutation relation depends on the value of the angle 
distribution at a point, which turns it of somewhat cumbersome 
handling.

By precise measurements on a light beam, a detailed test of the
uncertainty principle for angle and angular momentum has been 
recently demonstrated~\cite{Franke-Arnold:2004,Pegg:2005}. The 
idea is to pass the beam through an angular aperture and measure 
the resulting angular-momentum distribution~\cite{Leach:2002}.  In 
the same vein, we have presented experimental results~\cite{Hradil:2006} 
that strengthen the evidence that $\hat{E}$ furnishes a correct
description of angular phenomena. When a sensible periodic resolution 
measure (namely, the circular variance) is employed, the associated 
intelligent states should minimize two inequalities (one for the 
cosine and other for the sine), and both cannot be saturated 
simultaneously.  To bypass this drawback, we have looked at the 
more physically meaningful notion  of constrained intelligent 
states; that is, states that minimize  the uncertainty product 
for a given spread either in angle or in angular momentum. In 
fact, they prove to be   Mathieu  wave functions, which have 
been attracting great  interest in relation with nondiffracting 
optical fields~\cite{Gutierrez:2000,Gutierrez:2003,Bandres:2004}.

In this paper we go one step beyond and present an improved
experimental setup (that uses computer-controlled spatial light
modulators both at the state preparation and analyzing stages) to
verify in great detail the properties of these constrained intelligent
states. As a byproduct, we also bring out that $\hat E$ can be
associated with a feasible transformation (a fork-like hologram) 
that shifts the values of the angular momentum.  Our formulation 
paves thus the way for a full quantum processing of vortex beams 
and provides a bridge between the classical theory of singular 
optics and the realm of quantum optics.

The plan of this paper is as follows. In Sec.~\ref{sec:theory} 
we provide a comprehensive quantum treatment of angular variables,
including a discussion about the associated coherent and 
intelligent  states, as well as various suboptimal states.  
A feasible optical  realization of the system under study 
is provided in Sec.~\ref{sec:measurement}, with special 
emphasis  on the detection of the angular-momentum spectrum. 
In Sec.~\ref{sec:experiment} our setup is shown and experimental 
results are presented and discussed at length.  Finally, the 
summary of our achievements and suggestions for possible future 
upgrades are given in Sec.~\ref{sec:conclusions}.

\section{Theory}
\label{sec:theory}

\subsection{Quantum description of rotation angles}
\label{sec:angle}

We consider rotations of angle $\phi$ generated by the angular
momentum along the $z$ axis, which for the simplicity we shall 
denote henceforth as $L$. Classically, a point particle is 
necessarily located at a single value of the periodic coordinate 
$\phi$, defined within a chosen window, e.g., $[0, 2 \pi )$. The 
corresponding quantum wave function, however, is an object extended 
around the unit circle $\mathbb{S}_1$ and so can be directly affected 
by the nontrivial topology.

If we treat $\phi$ as a continuous variable, the Poisson 
bracket for the angle and the angular momentum is
\begin{equation}
  \{ \phi, L \} = 1 \, .
\end{equation}
Direct application of the correspondence between Poisson brackets 
and commutators, suggests the commutation relation (in units 
$\hbar = 1$)
\begin{equation}
  \label{ccr}
  [\hat{\phi}, \hat{L} ] = i \, .
\end{equation}
One is tempted to interpret $\hat{\phi}$ as multiplication by $\phi$, 
and then to represent $\hat{L}$ by the differential operator
\begin{equation}
  \label{Lz}
  \hat{L} = - i \frac{d}{d \phi}
\end{equation}
that verifies the fundamental relation (\ref{ccr}).  However, 
the use of this operator may entail many pitfalls for the unwary:
single-valuedness restricts the Hilbert space to the subspace of
$2\pi$-periodic functions, which, among other things, rules out 
the angle coordinate as a \textit{bona fide}
observable~\cite{Carruthers:1968,Emch:1972,Levy:1976}. A possible
solution, proposed by Judge and Lewis~\cite{Judge:1963}, is to 
modify the angle operator so that it corresponds to multiplication 
by $\phi$ plus a series of step functions that sharply change the 
angle by $2 \pi$ at appropriate points, which coincides with the 
classical Poisson bracket of $L$ and a periodic variable.

Many of these difficulties can be avoided by simply selecting instead
angular coordinates that are both periodic and continuous. However, a
single such quantity cannot uniquely specify a point on the circle
because periodicity implies extrema, which excludes a one-to-one
correspondence and hence is incompatible with uniqueness. Perhaps 
the simplest choice~\cite{Louisell:1963,Mackey:1963} is to adopt two
angular coordinates, such as, e. g., cosine and sine. In classical
mechanics this is indeed of a good definition, while in quantum
mechanics one would have to show that these variables, we shall 
denote by $\hat{C}$ and $\hat{S}$ to make no further assumptions 
about the angle itself, form a complete set of commuting operators.
One can concisely condense all this information using the complex
exponential of the angle $\hat{E} = \hat{C} + i \hat{S}$, which
satisfies the commutation relation
\begin{equation}
  \label{ELE} 
  [ \hat{E},  \hat{L} ] = \hat{E} \, .
\end{equation}
In mathematical terms, this defines the Lie algebra of the
two-dimensional Euclidean group E(2) that is precisely the canonical
symmetry group for the cylinder $\mathbb{S}_1 \times \mathbb{R}$
(i.e., the classical phase space of the system under study).

The action of $\hat{E}$ on the angular momentum basis is
\begin{equation}
  \label{E} 
  \hat{E} | m \rangle = | m -1 \rangle \, ,
\end{equation}
and, since the integer $m$ runs from $- \infty$ to $+ \infty$,
$\hat{E}$ is a unitary operator whose normalized eigenvectors are
\begin{equation}
  \label{phi_states} 
  |\phi \rangle = \frac{1}{\sqrt{2 \pi}}
  \sum_{m=- \infty}^\infty e^{i m \phi} | m \rangle \, .
\end{equation}
The intuitively expected relationship of a discrete Fourier pair
between angle and angular momentum is an immediate consequence of
Eqs.~(\ref{Lz}) and (\ref{phi_states}).  Indeed, denoting $\Psi_m =
\langle m | \Psi\rangle $ and $\Psi(\phi) = \langle \phi | \Psi
\rangle$, it holds
\begin{eqnarray}
  \label{Fourier}
  \Psi(\phi) & = &  \frac{1}{\sqrt{2 \pi}} 
  \sum_{m=- \infty}^\infty e^{-i m \phi} \,  \Psi_m \, , \nonumber \\
  & & \\
  \Psi_m  & = & \int_0^{2\pi} \! \! \! d\phi \,  \psi(\phi)
  e^{i m \phi} \, . \nonumber
\end{eqnarray}

There is an appealing physical interpretation beyond the definition 
of $\hat{E}$. Whereas in the case of (\ref{ccr}) one thinks in terms 
of complementarity between two measurable quantities, $\hat{E}$ 
primarily represents a transformation and (\ref{ELE}) may be 
interpreted as the complementarity between measurement and 
transformation. On the other hand, the action of $\hat{E}$ 
can be cast also in terms of measurement, since any unitary 
operator may be generated by an appropriate Hermitian generator. 
There is a twofold goal in the theory of angular momentum and 
its conjugate variable: to characterize them either as a 
transformation or as a measurement. Notice that the pioneering 
work in Refs.~\cite{Franke-Arnold:2004} and \cite{Pegg:2005} 
anticipated the former interpretation.

The role of $\hat{E}$ as a transformation is determined by
the action on the basis states~(\ref{E}). What deserves to be
explained is the possible measurement associated with $\hat{E}$.
Although the vectors $| \phi \rangle$ provide an adequate description
of angle, one must take into account that realistic measurements are
always imprecise. In particular, the measurement of the spectrum 
would require infinite energy.  In other words, the mathematical 
continuum of angles will be observed always with a finite resolution. 
In consequence, it could be interesting to  extend the previous 
formalism by including fuzzy, unsharp or noisy generalizations 
of the ideal description provided by $\hat{E}$.  To this end we 
shall use positive operator-valued measures (POVMs), that are a 
set of linear operators $\hat{\Lambda} (\phi)$ furnishing the 
correct probabilities in any measurement process through the
fundamental postulate that~\cite{Helstrom:1976}
\begin{equation}
  p (\phi) = \Tr [ \hat{\varrho} \, \hat{\Lambda} (\phi) ] \, ,
\end{equation}
for any state described by the density operator $\hat{\varrho}$.
Compatibility with the properties of ordinary probability imposes 
the requirements
\begin{equation}
  \label{condPOVM}
  \hat{\Lambda} (\phi) \ge 0 ,
  \quad
  \hat{\Lambda} (\phi) =  \hat{\Lambda}^\dagger (\phi) ,
  \quad
  \int_0^{2\pi} \! d\phi \; \hat{\Lambda} (\phi) = \hat{\openone} \, .
\end{equation}

In addition to these basic statistical conditions, some other
requisites must be imposed to ensure a meaningful description 
of angle as a canonically conjugate variable with respect 
$\hat{L}$.  We adopt the same axiomatic approach developed 
previously by Leonhardt \textit{et al}~\cite{Leonhardt:1995} 
for optical phase. First, we require the shifting property
\begin{equation}
  \label{CPOVM1}
  e^{i \phi^\prime \hat{L}} \, \hat{\Lambda} (\phi) \,
  e^{- i \phi^\prime \hat{L}}  =  \hat{\Lambda}
  (\phi + \phi^\prime) \, ,
\end{equation}
which reflects nothing but the basic feature that an angle shifter is
a angle-distribution shifter and imposes the following form for the
POVM~\cite{Luis:1998}
\begin{equation}
  \label{POVM1}
  \hat{\Lambda} (\phi)  = \frac{1}{2\pi}
  \sum_{m, m^\prime = - \infty}^\infty
  \lambda_{m, m^\prime} \, e^{i (m - m^\prime ) \phi} \,
  | m \rangle \langle m^\prime | \, .
\end{equation}
We must also take into account that a shift in $\hat{L}$ should not
change the phase distribution.  But a shift in $\hat{L}$ is generated
by $\hat{E}$ since, according to (\ref{E}), it shifts the angular
momentum distribution by one step. Therefore, we require as well
\begin{equation}
  \label{CPOVM2}
  \hat{E} \, \hat{\Lambda} (\phi) \, \hat{E}^\dagger  =  
  \hat{\Lambda} (\phi ) \, ,
\end{equation}
which, loosely speaking, is the physical translation of the fact that
angle is complementary to angular momentum. This imposes the additional 
constraint $\lambda_{m+1, m^\prime +1}= \lambda_{m, m^\prime}$,  and this 
means  that $\lambda_{m, m^\prime} = \lambda_{m - m^\prime}$. In  consequence, 
Eq.~(\ref{POVM1}) can be recast as
\begin{equation}
  \label{POVM2}
  \hat{\Lambda} (\phi)  = \frac{1}{2\pi}
  \sum_{l = - \infty}^{\infty}
  \lambda_{l}^\ast \, e^{- i l \phi} \,
  \hat{E}^{l} \, ,
\end{equation}
and the conditions (\ref{condPOVM}) are now
\begin{equation}
  | \lambda_l | \le 1 ,
  \qquad
  \lambda^\ast_{l} = \lambda_{- l} \, .
\end{equation}
Expressing $\hat{E}$ in terms of its eigenvectors, we finally arrive
at the more general form of the POVM describing the angle variable and
fulfilling the natural requirements (\ref{CPOVM1}) and (\ref{CPOVM2}):
\begin{equation}
  \label{POVM3}
  \hat{\Lambda} (\phi)  = \int_0^{2\pi} \! d\phi^\prime \;
  \mathcal{K} (\phi^\prime) \, |\phi + \phi^\prime \rangle
  \langle \phi + \phi^\prime | \, ,
\end{equation}
where
\begin{equation}
  \mathcal{K} (\phi)  = \frac{1}{2\pi}
  \sum_{l = - \infty}^{\infty}  \lambda_l \, e^{i l \phi} \, .
\end{equation}
The convolution (\ref{POVM3}) shows that this POVM effectively
represents a noisy version of the usual projection $| \phi \rangle
\langle \phi |$, and the kernel $\mathcal{K} (\phi)$ gives the
resolution provided by this POVM.

\subsection{Gaussian distributions on a circle}

Experience with quantum mechanics of simple systems, such as the free
particle and harmonic oscillator, suggests that Gaussian states can be
an important tool for a better understanding of the periodic motion on
a circle. Even so, it is remarkable that there is no clear concise
definition of the Gaussian distribution on a circle and one can only
find vague statements scattered through the literature.

We do not want to enter here into a mathematical treatment, but rather
we just try to grasp the properties that make the Gaussian distribution 
on the line to play such a key role in physics and that we are particularly 
keen on retaining when constructing its circular counterpart. We itemize 
the most relevant ones in our view:

\begin{enumerate}
\item The sum of many independent random variables tend to be
  distributed following a Gaussian distribution.
  
\item All marginal and conditional densities of a Gaussian are again
  Gaussians.

\item The Fourier transform of a Gaussian is also a Gaussian.

\item The Gaussian distribution maximizes the Shannon entropy for a
  fixed value of the variance.
\end{enumerate}

The first property (subject to a few general conditions) is the
central-limit theorem and explains the ubiquity of Gaussians in
physics: the distribution of the phenomenon under study does not have
to be Gaussian because its average will be.  The second and third ones
are responsible for the good properties than one assigns to Gaussian
states in quantum optics. Finally, the last condition bears on the
information-based approach to quantum theory, but strongly depends 
on the definition of entropy we adopt.

In statistics there are two distributions that have been somehow
suggested for having good properties on a circle, namely
\begin{eqnarray}
\label{pr_mises}
  p_{\kappa} ( \phi ) & = &   \frac{1}{2 \pi I_0 (2 \kappa)}
  \exp[2 \kappa \cos (\phi - \mu)] \, , \\
  & & \nonumber \\
\label{pr_wrapped}
 p_{\sigma} (\phi ) & = & \frac{1}{\sqrt{2 \pi} \sigma}
  \sum_{k = - \infty}^\infty \exp \left [ - \frac{1}{2} 
    \frac{(\phi - \mu + 2 \pi k)^2}{\sigma^2} \right ] \, ,
\end{eqnarray}
where $I_{n}$ denotes the modified Bessel function of first kind.
The first one is known as the von Mises distribution, while the 
second is the wrapped Gaussian.  By a trivial application of the 
Poisson summation formula we can express the latter as
\begin{equation}
  \label{theta3}
 p_{\sigma} (\phi ) = \frac{1}{2 \pi} \vartheta_{3} \left ( 
 \phi - \mu \biggl |  \frac{1}{e^{2 \sigma^{2}}} \right ) \, ,
\end{equation}
where 
\begin{equation}
  \label{eq:defJac}
  \vartheta_{3} (\zeta | q)  =
  \sum_{k=-\infty}^{\infty} q^{k^2} e^{2 i k \zeta}
\end{equation}
is the third Jacobi theta function~\cite{Mumford:1983,Abramowitz:1984}.
For both distributions $\mu$ represents the main direction, 
while $\sigma$ and $\kappa$ are parameters related to the 
concentration~\cite{Mardia:2000}.

From the previous checklist, the wrapped Gaussian satisfies 
properties 1 to 3, while the von Mises satisfies property 4 
when the variance  is replaced by its circular version (so it 
represents the minimally prejudiced angle distribution, given 
the information constraints~\cite{Barakat:1987}). Therefore, it 
is tempting to side with the former. Additionally, the Jacobi 
$\vartheta_{3}$ function is the solution of the diffusion equation 
on a circle with the initial state being a delta function, which 
is another way of defining a Gaussian wave function~\cite{Klimov:1997}. 
However, note that if we take such a route, Gaussian wave functions 
do not lead to Gaussian probability distributions anymore (because 
the square of a $\vartheta_3$ is not  a $\vartheta_3$ function), a 
limitation that does not apply to the von Mises.

In this respect, it is convenient to make a small detour into the
question of coherent states~\cite{Perelomov:1986} (recall that for the
harmonic oscillator they are precisely Gaussian wave packets). Possible 
definitions of coherent states for a particle on a circle have been 
outlined in the literature~\cite{Gonzalez:1998,Hall:2002}, but they 
are of very mathematical nature.  We prefer to adopt the ideas of 
Rembieli\'nski and coworkers~\cite{Kowalski:1996} and construct 
coherent states $|w\rangle$ as eigenstates of the operator
\begin{equation}
  \label{eq:defW}
  \hat{W} = e^{i(\hat{\phi} + i \hat{L})} = 
  e^{-\hat{L} + 1/2} \, \hat{E} \, ,
\end{equation}
so that
\begin{equation}
  \label{eq:coherst}
  \hat{W} | w \rangle = w | w \rangle \, ,
\end{equation}
where the complex number $w = e^{i \theta - \ell}$ parametrizes the
unit cylinder. Note in passing that 
\begin{eqnarray}
  \label{eq:cd}
  \hat{W} | m \rangle & = &  e^{m-1/2} | m - 1 \rangle \, ,
  \nonumber \\
  & & \\
  \hat{W}^\dagger | m \rangle & = & e^{m+1/2} | m + 1 \rangle \, ,
  \nonumber
\end{eqnarray}
with $[\hat{W}, \hat{W}^\dagger] = \sinh(1) e^{2 \hat{L}}$. The projection 
of the vector $| w \rangle$ onto the basis $|m \rangle$ gives then
\begin{equation}
  w_{m} = \langle m | w \rangle = w^{-m}e^{-m^2/2} \, ,
\end{equation}
while in the angular basis the corresponding expression is
\begin{equation}
  \label{eq:cohstaphi}
  w(\phi) = \frac{1}{\sqrt{2 \pi}} \vartheta_{3} \left ( 
\frac{1}{2}(\phi - \mu) \biggl |  \frac{1}{e^2} \right )  \, ,
\end{equation}
where $\mu = \theta + i \ell$. 
 
In consequence, we have found three families of states with
interesting properties: (i) states with von Mises probability
density, Eq.~\eqref{pr_mises}; (ii) states with wrapped Gaussian
\textit{probability} distribution, Eq.~\eqref{pr_wrapped}; and
(iii) coherent states with wrapped Gaussian  \textit{amplitude}
density, Eq.~\eqref{eq:cohstaphi}.  Nevertheless, leaving aside
fundamental reasons, for computational purposes these three 
families have very similar angular shapes and give almost
indistinguishable numerical results. Therefore, sometimes 
we will use von Mises states because of their simplicity 
and the possibility  of obtaining analytical results.

\subsection{Constrained intelligent states}
\label{sec:constrained}

Coherent states for the harmonic oscillator are also minimum
uncertainty wave packets. Given the analogy of $\hat{W}$ in
Eq.~(\ref{eq:cd}) with the standard annihilation operator, one 
is tempted to introduce quadrature-like combinations
\begin{equation}
  \label{eq:QP}
  \hat{Q} = \frac{1}{\sqrt{2}} (\hat{W} + \hat{W}^\dagger) \, ,
  \qquad
  \hat{P} = \frac{1}{\sqrt{2}i} (\hat{W} -  \hat{W}^\dagger) \, ,
\end{equation}
that satisfy the uncertainty principle
\begin{equation}
  \label{eq:uncQP}
  (\Delta \hat{Q})^2 \, (\Delta \hat{P})^2 \ge \frac{1}{4} 
  | \langle [\hat{Q}, \hat{P}]  \rangle |^2 \, ,
\end{equation}
where $(\Delta \hat{A})^2 = \langle \hat{A}^2 \rangle - \langle
\hat{A}^2 \rangle^2$ is the standard variance. A lengthy
calculation~\cite{Kastrup:2006} shows that the coherent states
(\ref{eq:cohstaphi}) obey (\ref{eq:uncQP}) as an equality and so 
they are indeed minimum packets for the variables (\ref{eq:QP}). 
In fact, they are also minimum for more intricate
uncertainties~\cite{Kowalski:2002}.  However, the problem is 
that, at difference of the harmonic oscillator, we do not have 
any clear operational prescription of how to measure the quadratures
(\ref{eq:QP}), so they give no real physical insight into the
statistical description of angle.

Let us then turn back to the general commutation relation (\ref{ELE}).
First, we observe that dealing with angle mean and variance in the
ordinary way has drawbacks. Consider, for example, a sharp angle
distribution localized at the origin and the same one shifted by
$\pi$.  Despite the fact that the physical information they convey is
the same, in the later case the variance is much bigger. Since angle 
is periodic but variance is not, it has little meaning to consider 
the angle measurement itself~\cite{Rao:1965}.

In circular statistics one usually calculates the moments of the 
exponential of the angle~\cite{Breitenberger:1985,Uffink:1990,
Bialynicki:1993,Forbes:2001}, that are referred to as circular 
moments and give rise, e. g., to a circular variance
\begin{equation}
  \sigma_{\phi}^2 = 1 - | \langle e^{i \phi} \rangle |^2 \, ,
\end{equation}
where
\begin{equation}
  \langle e^{i \phi} \rangle = \int_{0}^{2 \pi} d\phi \
 p(\phi) \ e^{i\phi} \, ,
\end{equation}
and $p(\phi)$ is the probability density.  It possesses all the 
good properties expected: it is periodic, the shifted distributions 
$P (\phi + \phi^\prime)$ are characterized by the same resolution, 
and for sharp angle distributions it coincides with the standard 
variance since $| \langle e^{i \phi} \rangle |^2 \simeq 1 + 
\langle \phi^2 \rangle$.  Moreover, this circular variance coincides 
with
\begin{equation}
  \label{defDE}
  (\Delta \hat{E})^2 = \langle \hat{E}^\dagger \hat{E} \rangle -
  \langle \hat{E}^\dagger \rangle \langle \hat{E} \rangle \, ,
\end{equation}
which is the natural extension of variance for unitary
operators~\cite{Levy:1976}. 

If we use (\ref{defDE}), the uncertainty relation associated
with (\ref{ELE}) reads
\begin{equation}
  \label{disp}
  ( \Delta \hat{E})^2  \, (\Delta \hat{L} )^2 \ge
  \frac{1}{4} [ 1 - ( \Delta \hat{E})^2 ] \, .
\end{equation}
Sometimes it probes convenient to express this in terms of the
corresponding Hermitian components $\hat{C}$ and $\hat{S}$. We have
\begin{equation} [\hat{C}, \hat{L} ] = i \hat{S} , \qquad [\hat{S},
  \hat{L} ] = - i \hat{C} \, ,
\end{equation}
while $[\hat{C}, \hat{S} ] = 0$, so that
\begin{eqnarray}
  \label{unCS}
  & ( \Delta \hat{C} )^2 (\Delta \hat{L} )^2 \ge
  \frac{1}{4} |\langle \hat{C} \rangle |^2 , & \nonumber \\
  & &  \\
  & ( \Delta \hat{S} )^2 ( \Delta \hat{L} )^2 \ge
  \frac{1}{4} |\langle \hat{S} \rangle |^2 \, . \nonumber &
\end{eqnarray}
Both inequalities depend on the choice of state used to evaluate
$\langle \hat{C} \rangle$ and $\langle \hat{S} \rangle$.  So
intelligent states need to be distinguished from minimum uncertainty
states: there are intelligent states for which the right-hand side of
Eq.~(\ref{unCS}) is not the obvious minimum value of 0. The condition
of intelligence for, say, the first of (\ref{unCS}), reads as
\begin{equation}
  (\hat{L} + i \kappa \hat{C} ) |\Psi \rangle =
  \mu |\Psi \rangle \, ,
\end{equation}
that once expressed in the angle representation can be immediately
solved to give
\begin{equation}
  \label{vMises}
  \Psi (\phi) = \frac{1}{\sqrt{2 \pi I_0 (2\kappa)}}
  \exp ( i \mu \phi + \kappa \sin \phi)  \, ,
\end{equation}
so that the associated probability is the von Mises distribution.  The
intelligent states for the the second equation in (\ref{unCS}) can be
worked out in the same way. However, it is not difficult to prove that
both inequalities cannot be saturated simultaneously~\cite{Bluhm:1995}.  
In other words, the fundamental relation (\ref{disp}) can never hold as 
an equality: it is exact, but is too weak.

%%%%%%%%%%%%%%%%%%%%%%%%%%%%%%%%%%%%%%%%%%%%%%%%%%%%%%%%%%%%%%%%%%%%%%%%%%%
\begin{figure}
\includegraphics[width=0.95\columnwidth]{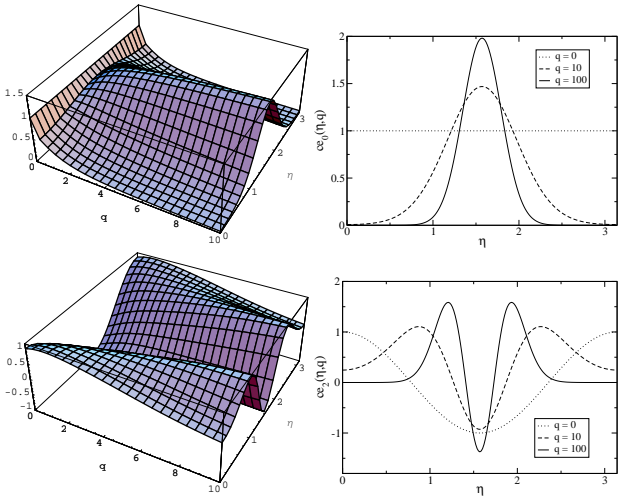}
\caption{Plot of the functions $\ce_0(\eta, q)$ (top) and
$\ce_2(\eta, q)$ (bottom). On the right, we show two-dimensional
sections of these functions for the values $q = 0, 1,2,3,4,$ and
100.
\label{fig:mathieu}}
\end{figure}
%%%%%%%%%%%%%%%%%%%%%%%%%%%%%%%%%%%%%%%%%%%%%%%%%%%%%%%%%%%%%%%%%%%%%%%%%%%

To get an attainable bound we look instead at normalized states that
minimize the uncertainty product $(\Delta \hat{E} )^2 \, (\Delta
\hat{L})^2$ either for a given $(\Delta \hat{E} )^2$ or for a given
$(\Delta \hat{L})^2$, which we call constrained intelligent states.
We use the method of undetermined multipliers, so  the linear
combination of variations leads to
\begin{equation}
  \label{eigcom}
  [ \hat{L}^2 +  p \, \hat{L} + ( q^\ast \hat{E} + q
  E^\dagger )/2 ] | \Psi \rangle = a | \Psi \rangle ,
\end{equation}
where $p$, $q$, and $a$ are Lagrange multipliers.  Working in
the angle representation, the change of variables $\exp(i p \phi)
\Psi (\phi)$ eliminates the linear term from (\ref{eigcom}). In
addition, we can take $q$ to be a real number, since this merely
introduces a global phase shift. We finally get
\begin{equation}
  \label{Mathieu}
  \frac{d^2 \Psi (\eta)}{d\eta^2}  + [ a - 2 q
  \cos (2 \eta) ] \ \Psi (\eta) = 0 ,
\end{equation}
where we have introduced the rescaled  variable $ \eta = \phi/2$, 
which has a domain $ 0 \le \eta < 2 \pi$ and plays the role of 
polar angle in elliptic coordinates. Equation (\ref{Mathieu}) is
precisely the standard form of the Mathieu equation, which has many
applications not only in optics, but also in other branches of modern
physics~\cite{McLachlan:1947}. An uncertainty relation of this type 
has been already investigated by Opatrn\'y~\cite{Opatrny:1995}. 
In our case, the only acceptable  Mathieu functions are those 
periodic with period of $\pi$ or $2 \pi$.  The values of $a$ in 
Eq.~(\ref{Mathieu}) that satisfy this condition are the eigenvalues. 
We have then two families of independent solutions, namely the angular 
Mathieu functions  $\ce_n ( \eta, q) $ and  $\se_n (\eta, q)$ with 
$n = 0, 1, 2, \ldots$,  which are usually known as the elliptic cosine 
and sine, respectively.  The parity of these functions is exactly the 
same as their trigonometric  counterparts; that is, $\ce_n (\eta, q)$ 
is even and $\se_n (\eta, q)$  is odd in $\eta$, while they have period 
$\pi$ when $n$ is even or  period $2 \pi$ when $n$ is odd.  To illustrate 
these behaviors, in  Fig.~\ref{fig:mathieu} we have plotted wave functions 
$\ce_n (\eta, q)$  of orders $n=0$ and $n=2$.

\begin{figure}
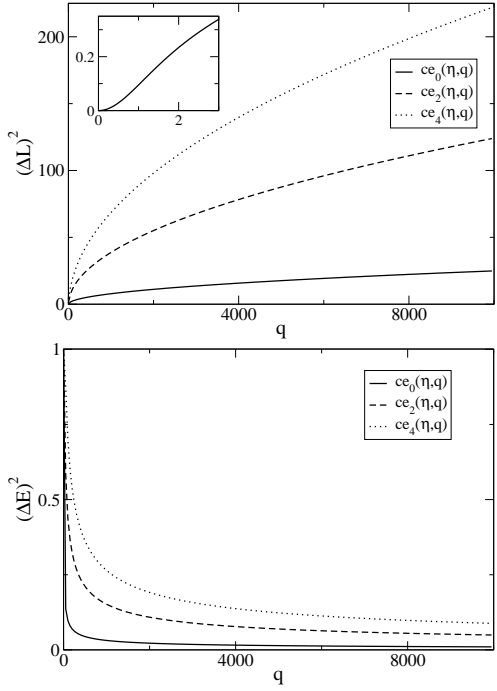

   \includegraphics[width=0.75\columnwidth]{Figure2a}\\
   \includegraphics[width=0.75\columnwidth]{Figure2b}
  \caption{Plot of $(\Delta \hat{L})_{2n}^2$ and $(\Delta \hat{E})_{2n}^2$ 
    for the three lowest order wavefunctions $\Psi_{2n} (\eta, q)$ in
    terms of the parameter $q$.
    \label{fig:dedl}}
\end{figure}

Since the $2\pi$ periodicity in $\phi$ requires $\pi$ periodicity in
$\eta$, the acceptable solutions for our eigenvalue problem are the
independent Mathieu functions $\ce_{2n} (\eta, q)$ and $\se_{2n}
(\eta, q)$, with $n = 0, 1, \ldots$. In what follows, we consider only
even solutions $\ce_{2n} (\eta, q)$, although the treatment can be
obviously extended to the odd ones with analogous results. We take
then
\begin{equation}
  \label{Matsol}
  \Psi_{2n} (\eta, q) = \sqrt{\frac{2}{\pi}}
  \ce_{2n} (\eta, q) \, ,
\end{equation}
where we have made use of the property
\begin{equation}
  \int_{0}^{2\pi} \ce_{m} (\eta, q) \ce_{n} (\eta, q) d\eta =
  \pi \delta_{mn} \, ,
\end{equation}
to normalize the wave function. Using (\ref{Matsol}) we have
\begin{eqnarray}
  (\Delta \hat{L})_{2n}^2 & = & \displaystyle
  \frac{1}{2\pi}  \int_{0}^{\pi}\! d\eta \,
  \ce_{2n}^{\prime \, 2} (\eta, q) \nonumber \\
  & =  & \frac{1}{4}[ A^{(2n)}_{2n}(q) - 2q \Theta_{2n} (q)] \, , 
  \nonumber \\
  &  & \\
  (\Delta \hat{E})_{2n}^2 & = &  1 -  \left | \frac{2}{\pi}
    \int_{0}^{\pi} \!d\eta \,
    \ce_{2n}^2 (\eta, q)  \, \cos (2\eta) \right |^2   \nonumber \\
  &  = &   1 - | \Theta_{2n} (q) |^2 \, . \nonumber 
  \label{dlde}
\end{eqnarray}
To obtain these analytical expressions we have expanded   
$\ce_{2n} (\eta, q)$  in Fourier series
\begin{equation}
  \ce_{2n} (\eta, q) = \sum_{k=0}^\infty A_{2k}^{(2n)} (q)
  \cos (2 k \eta) \, ,
\end{equation}
and integrate term by term, in such a way that
\begin{equation}
  \Theta_{2n} (q) =  A^{(2n)}_{0} (q) A^{(2n)}_{2} (q) +
  \sum_{k=0}^\infty A^{(2n)}_{2k} (q) A^{(2n)}_{2k +2} (q) \, .
\end{equation}
The coefficients $A_{2k}^{(2n)}$ determine the Fourier spectrum and
satisfy recurrence relations that can be efficiently computed by a
variety of methods~\cite{Frenkel:2001}. In Fig.~\ref{fig:dedl} we 
have plotted $(\Delta \hat{L})_{2n}^2$ and $(\Delta \hat{E})_{2n}^2$ 
as functions of the Lagrange multiplier $q$.

%%%%%%%%%%%%%%%%%%%%%%%%%%%%%%%%%%%%%%%%%%%%%%%%%%%%%%%%%%%%%%%%%%%%%%%%
%%%%%%%%%%%%%%%%%%%%%%%%%%%%%%%%%%%%%%%%%%%%%%%%%%%%%%%%%%%%%%%%%%%%%%%%%

\subsection{Intelligent states: asymptotic limits}
\label{sec:approx}

To better understand Fig.~\ref{fig:dedl} we first concentrate on 
the limit of small $q$ [i. e., large $(\Delta \hat{E})_{2n}^2$]. 
We expand $\ce_{2n} (\eta, q)$ in powers of $q$ and retain only 
linear terms~\cite{McLachlan:1947}
\begin{eqnarray}
  \label{ceqs}
  \ce_{0}(\eta, q) & = &\frac{1}{\sqrt{2}} 
  \left [ 1- \frac{q}{2}\cos (2 \eta) \right ] \, , 
  \nonumber \\
  \ce_{2}(\eta, q) & = & \cos (2 \eta) - q 
  \left [ \frac{\cos (4 \eta)}{12} - \frac{1}{4} \right ] \, , 
  \nonumber \\
  \ce_{2n}(\eta, q) & = & \cos (2 \eta ) - \frac{q}{4}
  \left \{ \frac{\cos[(2n+2)\eta]}{2n+1} \right. 
  \nonumber \\
  & - &\left. \frac{\cos[(2n-2)\eta]}{2n-1}\right\}, \quad n \ge 2.
\end{eqnarray}
This leads to
\begin{eqnarray}
  (\Delta \hat{L})_{2}^2 & = & 1 - \frac{5 q^2}{48} + O(q^4) \, ,
  \nonumber \\
  (\Delta \hat{E})_{2}^2 & = & 1- \frac{25q^2}{144} + O(q^4) \, , 
  \nonumber \\
  & & \\
  (\Delta \hat{L})_{2n}^2 & = &
  n^2 - \frac{q^2}{8(4n^2-1)} + O(q^4) \, , \nonumber \\
  (\Delta \hat{E})_{2n}^2 & = &  1 - \frac{q^2}{4 (4n^2 - 1)^2} + 
  O(q^4) \, , \qquad n \neq 1 
  \nonumber
\end{eqnarray}
showing a quadratic behavior which can be appreciated in
Fig.~\ref{fig:dedl}.  

In the opposite limit of large $q$ [small  $(\Delta \hat{E})_{2n}^2$] 
we take the approximation in terms of Hermite polynomials~\cite{Frenkel:2001}
\begin{equation}
  \label{Par}
  \ce_{2n}(\eta, q) \propto e^{-u^2/4} 
  H_{2n} \left(\frac{u}{\sqrt{2}}\right)  + O(q^{-1/2}) \, ,
\end{equation}
where $u = 2q^{1/4}\cos \eta$.  Apart from constant factors, the
states (\ref{Par}) look like the harmonic oscillator wave functions
and we can use them to evaluate analytically the variances. The final
expressions involve the modified Bessel functions $I_k(\sqrt{q})$, but
the crucial fact is that the following simple asymptotic expressions
hold:
\begin{eqnarray}
  \label{highq}
  (\Delta \hat{L})_{2n}^2 & = & \frac{(4n+1)}{4}\sqrt{q} + O(q^0) \, ,
  \nonumber \\
  & & \\
  (\Delta \hat{E})_{2n}^2 & = & \frac{4n+1}{\sqrt{q}} + O(q^{-1}) \, ,
  \nonumber
\end{eqnarray}
showing a square-root behavior that is also apparent from
Fig.~\ref{fig:dedl}.  The range of moderate values of $q$, where 
the transition between the quadratic (small $q$) and the square-root 
(large $q$) regions happens, is magnified in the inset.  According to
Eq.~\eqref{highq}, with increasing $q$ the uncertainty product
$(\Delta \hat{L})_{2n} (\Delta \hat{E})_{2n}$ approaches a constant
value depending exclusively on the mode index $n$; $\lim_{q \rightarrow
\infty}(\Delta \hat{L})_{2n} (\Delta \hat{E})_{2n}= (4n+1)/2$. 
These asymptotic limits, confirmed in Fig.~\ref{fig:dedl}, identify 
the fundamental mode $n=0$ as the minimum uncertainty state for all 
the values of the parameter $q$. Henceforth, we always refer to the 
fundamental Mathieu mode, unless the mode index is explicitly given.

Finally, let us note that, from Eq.~\eqref{ceqs}, it follows immediately
that when $q \rightarrow 0$ the probability distribution for this
fundamental mode is $p(\phi) \propto [1- q \cos(\phi)/2]^2 
\simeq \exp(-q \cos\phi)$; while when $q \rightarrow \infty$, according
to Eq.~\eqref{Par},  we have $H_0(u/\sqrt{2})=1$ and $p(\phi) \propto
\exp[-2\sqrt{q}\cos^2(\phi/2)]\propto \exp(-\sqrt{q}\cos\phi)$. We
therefore get the interesting result that
\begin{equation}
  \label{mises}
  p (\phi ) \propto | \ce_0 (\eta, q) |^2 \simeq
  \left \{
    \begin{array}{ll}
      e^{- q \cos \phi}, \quad &   q \rightarrow 0 ,\\
      & \\
      e^{- \sqrt{q} \cos \phi} ,  \quad &   q \rightarrow \infty ,
    \end{array}
  \right .
\end{equation}
and hence the optimal states with very sharp and nearly flat angular
profiles attain the von Mises shape.

\subsection{Suboptimal states}
\label{sec:subopt}

Up to now we have investigated extremal states that will be used in
the experiments as an ultimate calibration to asses the performance of
our setup.  Here, we compare these extremal states with suboptimal
ones. There are a plenty of possible candidates for that: we will
select a few examples that can be easily prepared and intuitively grasp various features of ``a well localized angle''.

The wedge structure is our first representative. The aperture function 
possesses  sharp edges and may be defined in the angle representation as
\begin{equation}
  \label{wedge}
  \Psi( \phi ) =
  \left \{
    \begin{array}{lcl}
      1 / \sqrt{\alpha} \, , \qquad & 
      | \phi |\le \alpha/2 \, , \\[5pt]
      0 \, , &  | \phi | > \alpha/2 \, ,
    \end{array}
\right.
\end{equation}
$\alpha $ being the opening angle of the wedge. The probability
distribution of angular momentum $p_m= |\Psi_m|^2$ can be calculated
using Eq.~\eqref{Fourier} and one finds 
\begin{equation}
  \label{wedge-distribution}  
  \begin{split}
    (\Delta \hat{L})^2 & \rightarrow \infty \, , \\
    p_m &= \frac{\alpha}{2\pi} \frac{\sin^2(m \alpha/2)}{(m\alpha/2)^2} \, ,
    \qquad \text{[wedge]}\\
    (\Delta \hat{E})^2 & = 1 - \frac{4}{\alpha^2} \sin^2(\alpha/2) \, .
  \end{split}
\end{equation}
Similarly to the Fraunhofer diffraction pattern observed behind a
rectangular slit, the variance $(\Delta \hat{L})^2$ is infinite due to
the heavy tails of the $\mathrm{sinc}$ distribution and thus the
angle-angular momentum uncertainty relation is trivially satisfied for
(\ref{wedge}).  In spite of this divergence, the experimenter cannot
establish this simple fact from a finite data set: in general, the
sampled angular-momentum uncertainty will grow with the size of data
acquired.  Most of the detected $m$ fall within the central peak $|m|
\le 2\pi/\alpha$ of the distribution, which tends to regularize the
unbounded uncertainty product.

As another candidate for a simple single-peaked angular distribution
we choose the state
\begin{equation}
  \label{cosine}
  \Psi(\phi) =\left\{
    \begin{array}{lcl} \displaystyle
      \sqrt{\frac{2}{\pi\alpha}} \cos( \phi/\alpha ), \qquad &
      | \phi | \le \pi \alpha/ 2 \, ,\\[8pt]
      0, & | \phi | > \pi \alpha / 2 \, ,
    \end{array}
  \right.
\end{equation}
that is, the positive cosine half-wave stretched to the interval $-
\pi \alpha/2 \le \phi \le \pi \alpha/2$, so that $\alpha \le 2$. The
following discussion is valid even for $\alpha > 2$ provided the 
cosine half-wave, which now spans an interval of width larger 
than $2\pi$, is wrapped onto the unit circle. Since the delimiting
aperture has no sharp edges, one can expect more regular
results. A straightforward calculation yields
\begin{equation}
  \label{cosinevar}
  \begin{split}
    (\Delta \hat{L})^2 & = 1/\alpha^2 \, , \\
    p_m &=\frac{4\alpha \cos^2(\pi m \alpha/2)}
    {\pi^2(m^2\alpha^2-1)^2} \, , \qquad \text{[cosine]}\\
    (\Delta \hat{E})^2 & = 1 - \frac{64 \sin^2(\pi \alpha/2)}
    {\pi^2\alpha^2(\alpha^2-4)^2} \, .
  \end{split}
\end{equation}
At first sight, the angular-momentum distribution in
Eq.~\eqref{cosinevar} has a sinc-like shape with infinitely 
many  side lobes, which strongly resembles the one in
Eq.~\eqref{wedge-distribution}. However, here the higher-order
contributions to the angular-momentum variance are negligible 
and  both uncertainties appear to be finite.  Notice that 
the parameter $\alpha$ has now a very simple physical meaning: 
it  is inversely proportional to the angular-momentum uncertainty.

Next, we consider the von Mises wave function
\begin{equation}
  \label{centered_mises}
  \Psi( \phi ) = \frac{1}{\sqrt{2\pi I_0(1/\alpha)}} 
  e^{\cos\phi/(2\alpha)},
\end{equation}
$\alpha$ being a monotonic function of the angular width. According to
Eq.~\eqref{mises}, the relevant uncertainties,
\begin{equation}
  \begin{split}
    (\Delta \hat{L})^2 &= \frac{I_1(1/\alpha)} 
    {4 \alpha I_0 (1/\alpha)} \, ,   \\
    p_m &=\frac{I^2_m(1/2\alpha)}{I_0(1/\alpha)} \, ,
    \qquad  \text{[von Mises]} \\
    (\Delta \hat{E})^2 & = 1 -\frac{I^2_1 (1/\alpha)}
    {I^2_0(1/\alpha)} \, ,
  \end{split}
\end{equation}
can be  also taken (with an appropriate fitting of the parameter  
$\alpha$) as excellent approximations to the uncertainties of 
the optimal Mathieu states, especially in the regions of small 
and large variances.

%%%%%%%%%%%%%%%%%%%%%%%%%%%%%%%%%%%%%%%%%%%%%%%%%%%%%%%%%%%%%%%%%%%%%%%%%%%
\begin{figure}
  \centerline{\includegraphics[width=.90\columnwidth]{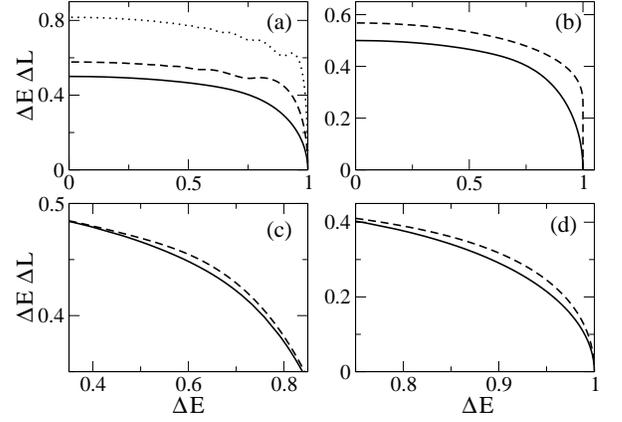}}
\caption{Theoretical uncertainty products for angle and angular-momentum 
variables calculated for (a) wedge angle distribution, (b) cosine
distribution; (c) von Mises distribution, and (d) truncated normal  
distribution.  In all panels the solid line denotes the optimal
uncertainty product generated by the intelligent Mathieu wave 
function. In panel (a) the broken and dotted lines correspond 
to $p_m$  truncated at the first and second minimum, respectively. 
\label{fig:comparison}}
\end{figure}
%%%%%%%%%%%%%%%%%%%%%%%%%%%%%%%%%%%%%%%%%%%%%%%%%%%%%%%%%%%%%%%%%%%%%%%%%%%%

Finally, we evaluate the uncertainties of the truncated Gaussian
\begin{equation}
\Psi ( \phi )= [\pi \, \mathrm{erf}^2(\pi\alpha)/\alpha^2]^{-1/4}
e^{-\alpha^2 \phi^2/2} \, ,
\end{equation}
which minimizes the uncertainty product for angular momentum and
angle, the latter in the sense of ordinary
variance~\cite{Franke-Arnold:2004,Pegg:2005}.  Such states become
suboptimal when the variance is replaced by a periodic measure,
such as the circular variance  advocated in this paper: 
\begin{equation}
  \begin{split}
    (\Delta \hat{L})^2 & = \frac{\alpha^2}{2} \left [ 1 - \frac{2
        \sqrt{\pi} \alpha e^{-\pi^2 \alpha^2}}
      {\mathrm{erf} (\pi\alpha)} \right ] \, , \\
    p_m & = \frac{ e^{-m^2/\alpha^2} \left \{ \mathrm{Re} \left [
          \mathrm{erf} \left ( \frac{\pi\alpha^2+i m}{\sqrt{2} \alpha}
          \right) \right] \right \}^2}{\sqrt{\pi}\alpha
       \,\mathrm{erf}^2(\pi\alpha)} \, ,
    \quad \text{[truncated]}\\
    (\Delta \hat{E})^2 &= 1 - \frac{e^{-1/(2 \alpha^2)} \left \{ 
        \mathrm{Re} \left [ \mathrm{erf} \left ( \frac{2\pi\alpha^2 + i}
            {2\alpha} \right ) \right] \right \}^2} {\mathrm{erf}^2(\pi\alpha)} \, ,
  \end{split}
\end{equation}
where $\mathrm{erf}(z)$ is the error function.

Figure~\ref{fig:comparison} shows the comparison of these four states
with the optimal Mathieu one.  As it has been already mentioned, the
measured uncertainty product for the wedge distribution grows with the
size of the data acquired, since more and more side maxima are sampled
[see panel (a)].  On the other hand, the uncertainty product of the
cosine distribution in panel (b) is well defined and lies well above
the quantum limit given by the Mathieu profile.  In agreement with the
asymptotic analysis of the previous section, the uncertainty product
for the von Mises angular distribution falls very close to the optimal
curve: only at intermediate angular spreads $\Delta \hat{E}$ we see a
significant deviation from the standard quantum limit, while for the
truncated Gaussian states the deviation is larger and shifts toward
higher values of $\Delta \hat{E}$.  The point is whether the currently
available measurements have sufficient resolution to discriminate
between the optimal and suboptimal states mentioned above.  This
question is addressed in the next two sections.

\section{Measurement of the orbital angular momentum}
\label{sec:measurement}

\subsection{Single vortex beam}
\label{sec:vortex}

The theory presented thus far can be applied to a variety of physical
systems.  Here, we consider a particularly appealing realization of a
planar rotator in terms of optical beams.

Light beams can carry angular momentum, which comprises spin and
orbital components that are associated with polarization and
helical-phase fronts, respectively. In general, the spin and orbital
contributions cannot be considered separately, but in the paraxial
approximation both contributions can be measured and manipulated
independently. We emphasize that this OAM manifests at the macroscopic
and single-photon levels and therefore paraxial quantum optics is the
most convenient context in which to treat the OAM of light as a
quantum resource.
 
In consequence, we can leave aside the spin part and consider the
simplest scalar monochromatic beam carrying OAM: this is precisely a
vortex beam; i.e., a beam whose phase varies in a corkscrew-like
manner along the direction of propagation. The corresponding spatial
amplitude can be written as
\begin{eqnarray}
  U (\mathbf{r}) = u (\mathbf{r}) 
  \exp (i m \phi ) \, ,
  \label{gvort}
\end{eqnarray}
where we have assumed that the beam propagates dominantly along the
$z$ axis, so we have cylindrical symmetry.  According to the
representation in Eq.~(\ref{Lz}), (\ref{gvort}) is an eigenstate of
$\hat{L}$ with eigenvalue $m$, which is also known as the topological
charge (or helicity) of the vortex.  To check this interpretation,
note that, the OAM density is also dominantly along the $z$ axis and
is given by $l = r S_{\phi}/c^{2}$ , where $S_{\phi}$ is the azimuthal
component of the Poynting vector. In a scalar theory, the
time-averaged Poynting vector can be calculated computed as
$\mathbf{S} = is_{0}\omega(U^\ast \nabla U - U \nabla U^\ast)$, where
$s_{0}$ is a constant (with units of m s).  The density of the OAM of
the vortex beam (\ref{gvort}) then depends on its intensity
$I=|U|^{2}$ and wavefront helicity and can be expressed in terms of
its power $P$ as
\begin{eqnarray}
  L = \frac{2\omega s_{0} m P}{c^{2}}.
  \label{totOAM}
\end{eqnarray}
If we divide now by the total energy density of the field we finally
get that the OAM per photon can be interpreted precisely as the
topological charge $m$. In this way, light beams prepared in 
OAM eigenstates can be used in quantum optics experiments in
the same way as qudits.

\subsection{Principle of  the measuring method}
\label{sec:detection}

\begin{figure*}
\includegraphics[width=1.90\columnwidth]{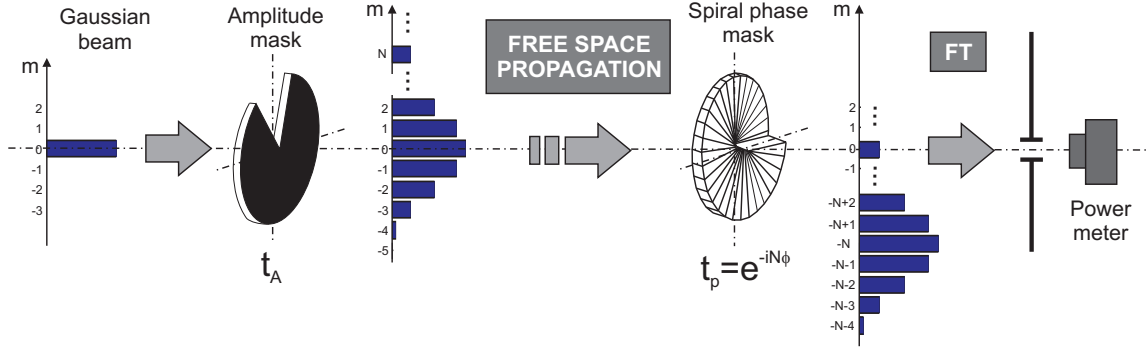}
\caption{Principle of the measurement of the angular momentum 
spectrum of the angular restricted field.
\label{fig:pmm}}
\end{figure*}

A general scheme of our experimental method is sketched in
Fig.~\ref{fig:pmm}. A collimated Gaussian beam with complex amplitude
$U_{G}$ illuminates an amplitude mask (with transmission coefficient
$t_{A}$) performing an angular limitation of the beam.  Immediately
behind the mask, the beam transverse profile has a cake-slice shape
given by $U_{A} = t_{A} U_{G}$. According to Eq.~\eqref{Fourier}, the
field azimuthal amplitude distribution results in a spread of the
spectrum composed of vortex components with different topological 
charges and amplitudes. The beam propagates toward a spiral phase 
mask [with transmission $t_{P}=\exp(-i N \phi)$] introducing a 
helicity $-N$. Behind the phase mask, the transmitted field is 
Fourier analyzed, so its spatial spectrum $\overline{U}_{F}$ can 
be obtained as
\begin{equation}
  \overline{U}_{F} = \mathcal{F}[ (t_{A} U_{G} \ast h) t_{P} ] \, ,
\end{equation}
where $\mathcal{F}$ denotes the Fourier transform, $h$ is the impulse
response function of free-space propagation between the amplitude
and phase masks and $\ast$ is the convolution product.  The detected
power can be thus interpreted as the spectral intensity collected at
the power meter placed at the back focal plane of the lens used for 
the optical implementation of the Fourier transform.

%%%%%%%%%%%%%%%%%%%%%%%%%%%%%%%%%%%%%%%%%%%%%%%%%%%%%%%%%%%%%%%%%%%%%%
%%%%%%%%%%%%%%%%%%%%%%%%%%%%%%%%%%%%%%%%%%%%%%%%%%%%%%%%%%%%%%%%%%%%%%

As we show below, provided the aperture radius of the power meter 
is suitably chosen, the measured power can be used for estimating 
the OAM of the vortex mode with topological charge $N$ in the field 
behind the amplitude mask.  This procedure can be repeated with 
spiral phase masks of different topological charges yielding the 
vortex distribution (i.e., angular-momentum spectrum) of the  
prepared angular-restricted beam.  Several experimental 
realizations of this idea have been proposed and
realized~\cite{Leach:2002,Pegg:2005,Hradil:2006}, differing in
technical details and data analysis~\cite{Vasnetsov:2003,Molina:2007b}.

To put this in quantitative terms, let us first introduce a mode 
decomposition of the amplitude mask
\begin{equation}
  t_{A}=\sum_{m=-\infty}^{\infty} a_{m}\exp(i m \phi) \, ,
\label{IFS}
\end{equation}
where $a_{m}$ are Fourier coefficients. We assume that the waist of
the Gaussian beam (of width $w_{0}$) is placed exactly at the mask
plane. The transmitted field propagates through free space, so the
complex amplitude $U_{I}$ of the field impinging on the spiral phase
mask can be written in the form
\begin{equation}
  U_{I}(\mathbf{r}) = \sum_{m=-\infty}^{\infty}a_{m} u_{m}(r,z) \exp(im\phi),
  \label{GFS}
\end{equation}
where we have used cylindrical coordinates $(r, \phi, z)$ and
\begin{eqnarray}
  & u_{m}(r,z)  =  2 \pi u_{0} i^{m} h_{0}
  \exp\left(-\frac{ikr^{2}}{2z}\right)A_{m}(r,z) \, , & \nonumber \\
  & \displaystyle 
  A_{m}(r,z)  =  \int_{0}^{\infty}\exp(-\alpha r^{\prime}{}^{2})
  J_{m} (\beta rr^\prime )r^\prime dr^\prime \, , & \label{integral} \\
  & \displaystyle 
  h_{0} = \frac{i}{\lambda z} \, ,  \qquad
  \alpha = \frac{1}{w_{0}^{2}} + \frac{ik}{2z} \, , \qquad
  \beta = \frac{k}{z} \, . & 
  \nonumber
\end{eqnarray}
Here $J_{m}$ and $u_{0}$ denote the $m$-th order Bessel function of
the first kind and a constant amplitude of the Gaussian beam,
respectively, and $z$ is the distance between amplitude and phase
masks. The integration in (\ref{integral}) can be carried out and 
results  in
\begin{eqnarray}
  A_{m}(r,z)&=&r\frac{Q}{\beta}\sqrt{\frac{\pi}{\alpha}}\exp(-Qr^{2})
  \nonumber \\
  & \times & [I_{\frac{1}{2}(m-1)}(Qr^{2}) - 
  I_{\frac{1}{2} (m+1)}(Qr^{2}) ], 
\end{eqnarray}
where $ Q = \beta^{2}/(8 \alpha)$.

After transmission through the spiral phase mask, the Fourier transform
of the field is performed optically and  the spatial distribution 
at the back focal plane of the Fourier lens can be represented by the
complex amplitude $\overline{U}_{F}$ given by
\begin{eqnarray}
  \label{casignal}
  \overline{U}_{F}(\nu,\psi) & = &
  a_{N}[ \overline{u}_{N}(\nu) - \overline{v}_{N}(\nu)]
  \nonumber \\
  & + &\sum_{m=-\infty}^{\infty}a_{m} \overline{v}_{m}(\nu) \exp[i(m-N)\psi] \, ,
\end{eqnarray}
where
\begin{eqnarray}
  \overline{u}_{m}(\nu)&=& 2 \pi \int_{0}^{\infty} \! \! \! 
  u_{m}(r,z)J_{0}(2\pi\nu r) r dr \, ,
  \nonumber \\
  & & \\
  \overline{v}_{m}(\nu)&=&2\pi i^{(m-N)}\int_{0}^{\infty} \! \! \!
  u_{m}(r,z)J_{m-N} (2\pi\nu r) r dr \, .
  \nonumber
\end{eqnarray}
Here $(\nu, \psi)$ are polar coordinates in the transverse Fourier
plane $(x_{F},y_{F})$, defined as $\nu = \sqrt{x_{F}^{2}+ y_{F}^{2}}/
(\lambda f^\prime)$ and $\psi = \arctan (y_F/x_F)$,  $f^\prime$ being 
the lens focal length.  The power captured by the circular aperture 
of the meter placed at the focal plane of the lens is given by
\begin{equation}
  P = \int_{0}^{\nu_{0}}\int_{0}^{2\pi}|\overline{U}_{F}(\nu,\psi)|^{2}
  \nu d\nu d\psi \, ,
  \label{defpower}
\end{equation}
where $\nu_{0}=R/(\lambda f')$ and $R$ denotes the aperture
radius. Substituting (\ref{casignal}) into (\ref{defpower}), the 
detected power can be expressed as
\begin{equation}
  P=P_{N}+P_{C} \, ,
\end{equation}
where
\begin{equation}
  \label{signal-noise}
  \begin{split}
    P_{N}& =2 \pi |a_{N}|^{2}\int_{0}^{\nu_{0}}|\overline{u}_{N}|^{2}\nu d\nu,\\
    P_{C}& =2 \pi \sum_{\substack{m=-\infty \\  m \neq N}}^{\infty}
    \!\!\!\!|a_{m}|^{2}\int_{0}^{\nu_{0}}|\overline{v}_{m}|^{2}\nu
    d\nu \, .
  \end{split}
\end{equation}
This power appears as composed of two terms: the first term, $P_{N}$,
represents the power carried by the vortex mode of topological charge 
$N$. The second term, $P_{C}$, is a crosstalk that represents disturbing 
contributions of the remaining vortex components and it can be reduced 
by a convenient choice of the aperture radius $R$ of the power meter.  
This possibility follows directly from Eq.~(\ref{casignal}), since
$|\overline{u}_{N}(0)|^{2}  \neq 0$ and has its maximum in the middle
of the receiving aperture, while $|\overline{v}_{m}(0)|^{2}=0$ for 
$m \neq N$.  The spectral intensity of the vortex whose power is 
measured creates a sharply  peaked bright spot at the Fourier plane,
meanwhile, the spectral intensities contributing to the crosstalk 
power have an annular form. These different spatial shapes enable 
an optimal choice of the radius  of the receiving aperture. In this 
case, both the power lost of the measured vortex and the influence 
of the crosstalk are minimized.

\section{Experimental results}
\label{sec:experiment}

To verify our theory, the angle-angular momentum uncertainty products 
were experimentally measured on various light beams.  Given the small 
difference between  the optimal Mathieu beams and other suboptimal 
single-peaked angular distributions, such a measurement is also an 
indicator of the resolution attainable with the present commercially 
available technology.

%%%%%%%%%%%%%%%%%%%%%%%%%%%%%%%%%%%%%%%%%%%%%%%%%%%%%%%%%%%%%%%%%%%%%%%%
\begin{figure}
  \includegraphics[width=0.99\columnwidth]{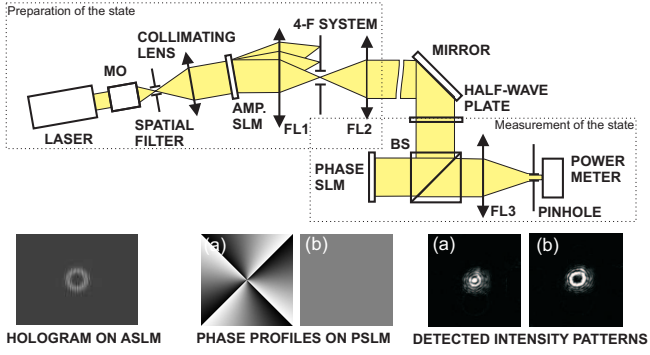}
  \caption{Experimental setup for the generation of
    beams with an arbitrary transverse profile and subsequent
    detection of their angular momentum spectrum.}
 \label{fig5}
\end{figure}
%%%%%%%%%%%%%%%%%%%%%%%%%%%%%%%%%%%%%%%%%%%%%%%%%%%%%%%%%%%%%%%%%%%%%%%

Figure~\ref{fig5} shows our setup. The beam generated by the
laser (Verdi V2: 532 nm, 20~mW) is spatially filtered, expanded and
collimated by the lens and impinges on the hologram generated by the
amplitude spatial light modulator (SLM) (CRL Opto, $1024 \times 768$
pixels).  The bitmap of the hologram is computed as an interference
pattern of the tested state (with the desired angular amplitude
distribution) and an inclined reference plane wave.  After
illuminating the hologram with the collimated beam, the Fourier
spectrum of the transmitted beam is localized at the back focal plane
of the first Fourier lens FL$_1$. It consists of three diffraction
orders $(-1, 0, +1)$. The undesired 0 and $-1$ orders are removed by a
spatial filter. After inverse Fourier transformation, performed by the
second Fourier lens FL$_2$, a collimated beam with the required
complex amplitude profile $U_A=t_A U_G$ is obtained. This completes
the state preparation.

The analysis begins by reflecting the prepared field $U_{A}$ at a phase
SLM (Boulder, $512 \times 512$ pixels), whose reflectivity is
proportional to $t_P\propto e^{-i N \phi}$.  As it has been discussed 
in the previous section, after the Fourier transformation of the
reflected field, the spectral component whose helicity was eliminated 
by the phase SLM gives rise to a bright spot (Fig.~\ref{fig5}a), 
while the other components have an annular intensity  distribution 
(Fig.~\ref{fig5}b). The vortex components of the spiral spectrum 
can be subsequently selected by the phase SLM and their OAM determined 
by a power measurement performed with an  optimal aperture size of the 
power meter. To suppress crosstalks, the calibrating response functions 
were acquired for each phase mask.

%%%%%%%%%%%%%%%%%%%%%%%%%%%%%%%%%%%%%%%%%%%%%%%%%%%%%%%%%%%%%%%%%%%%%%
\begin{figure}
  \includegraphics[width=0.7\columnwidth]{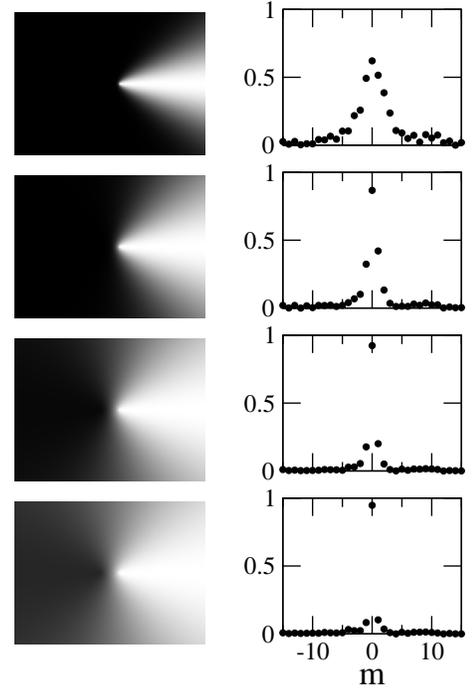}
  \caption{Preparation and measurement of Mathieu states having
    circular variances (from top to bottom) $\Delta \hat{E} = 0.31,
    0.54, 0.79$, and 0.91.  Left: computed intensity distribution;
    right: measured angular-momentum spectrum (in arbitrary units).}
  \label{fig:data}
\end{figure}
%%%%%%%%%%%%%%%%%%%%%%%%%%%%%%%%%%%%%%%%%%%%%%%%%%%%%%%%%%%%%%%%%%%%%

After the setup was carefully aligned using Laguerre-Gauss beams,
transverse amplitude distributions of different shapes and angular
variances were generated.  Each beam was then scanned for values of
helicities in the range of $m \in [-15,15]$. A typical transverse
intensity profile and the corresponding measured raw data are shown
in Fig.~\ref{fig:data}.

%%%%%%%%%%%%%%%%%%%%%%%%%%%%%%%%%%%%%%%%%%%%%%%%%%%%%%%%%%%%%%%%%%%%
\begin{figure}
  \includegraphics[width=0.8\columnwidth]{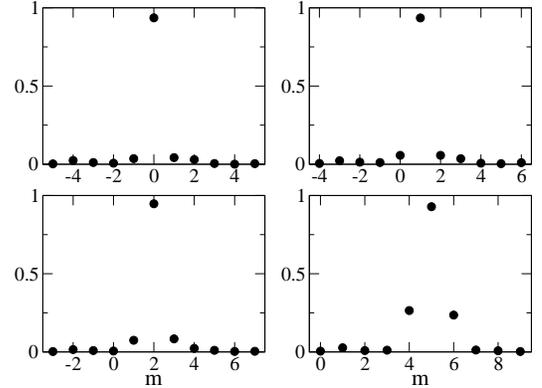}
\caption{Measured response functions of our detection scheme
for pure vortex modes of helicities $m= 0,1,2$, and 5.}
\label{fig:crosstalk}
\end{figure}
%%%%%%%%%%%%%%%%%%%%%%%%%%%%%%%%%%%%%%%%%%%%%%%%%%%%%%%%%%%%%%%%%%%%

In addition, the response functions were measured for pure vortex
modes (see Fig.~\ref{fig:crosstalk}).  For an ideal detection, the
pure vortex mode with the topological charge $N$ should have a
$\delta$-like angular momentum spectrum with a sharp peak at $m=N$. 
Any real detection scheme suffers from crosstalks between modes, 
which tends to broaden the measured spectra in Fig.~\ref{fig:data}; 
this effect becomes more pronounced for larger helicities ($m>4$).  
It can be seen, from comparing Figs.~\ref{fig:data} and 
\ref{fig:crosstalk}, that the reliability of the measured 
spectrum decreases from the center ($m=0$) to the edges and 
that beams of smaller variances have broader angular spectra 
and {\it vice versa}. Hence the reliability of experimentally 
determined uncertainty products is expected to increase
with variance.

In the next step, the acquired response functions were used to
increase the resolution of our detection scheme.  Since the measured
spectra could be considered to be convolutions of true angular spectra
and known response functions, we could apply an inverse transformation
to minimize the effect of crosstalks.  Then, the angular-momentum
variances were estimated by fitting the deconvoluted data to the
theoretically-calculated distributions.  The family of distributions
used for fitting experimental data was parametrized by an overall
normalization factor and a parameter characterizing the angular width
of the corresponding state.  For example, the fitting procedure
applied to data measured on a Mathieu beam yielded the value of
parameter $q$, which was then used to determine the variance of
angular momentum via Eq.~\eqref{dlde}.
%%%%%%%%%%%%%%%%%%%%%%%%%%%%%%%%%%%%%%%%%%%%%%%%%%%%%%%%%%%%%%%%%%%%
\begin{figure}
  \includegraphics[width=0.75\columnwidth]{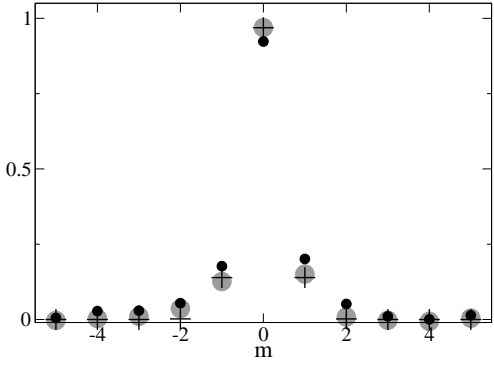}
\caption{Analysis of measured angular momentum spectrum 
of a Mathieu beam of $\Delta \hat{E} = 0.79$. Raw data 
(small black circles), deconvoluted data (large gray circles) 
and best fit with a theoretical distribution (+ symbols) are shown.
\label{fig:fit}}
\end{figure}
%%%%%%%%%%%%%%%%%%%%%%%%%%%%%%%%%%%%%%%%%%%%%%%%%%%%%%%%%%%%%%%%%%%%
This stage of analysis is illustrated in Fig.~\ref{fig:fit}.
Besides getting $(\Delta \hat{L}^2)$ the quality of the best fit was
quantified enabling to place error-bars on the resulting uncertainty
products.
 
%%%%%%%%%%%%%%%%%%%%%%%%%%%%%%%%%%%%%%%%%%%%%%%%%%%%%%%%%%%%%%%%%%%%%
\begin{figure}
  \includegraphics[width=\columnwidth]{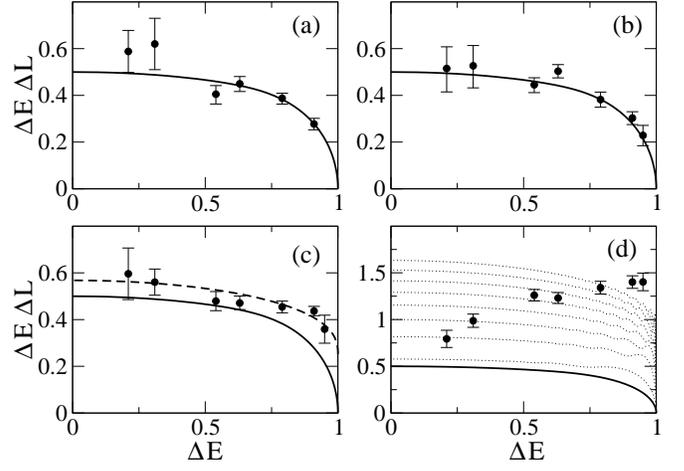}
\caption{Experimentally determined uncertainty products for
angle and angular momentum. The following angular distributions
of beam amplitude in the transverse plane were measured: 
(a) Mathieu distribution of Eq.~\eqref{Matsol}, (b) von Mises distribution 
of Eq.~\eqref{centered_mises}, 
(c) cosine distribution of Eq.~\eqref{cosine}, and
(d) wedge distribution of Eq.~\eqref{wedge}.
Experimentally obtained uncertainty products are 
denoted by circles.
For comparison, theoretical uncertainty products 
of the optimal Mathieu angular distribution (solid line),
cosine angular distribution (broken line), and  
wedge angular distributions whose angular momentum
spectra have been truncated at the first, second, etc. minimum 
(dotted lines) are also shown.
\label{fig:experiment}}
\end{figure}
%%%%%%%%%%%%%%%%%%%%%%%%%%%%%%%%%%%%%%%%%%%%%%%%%%%%%%%%%%%%%%%%%%%%%%

Experimental results are summarized in Fig.~\ref{fig:experiment}.
Given the resolution of the setup (indicated by error bars), the
obtained uncertainty products fit quite well the theoretical
predictions.  As anticipated, the resolution is not uniform and gets
better in the region of large variances.

The inspection of the upper panels of Fig.~\ref{fig:experiment} shows
that the measurements of the optimal Mathieu and von Mises beams yield
very similar results. This could be expected, since the difference
between the uncertainty products of these two beams (see
Fig.~\ref{fig:comparison}c) is below the resolution of the present
setup.  The cosine and wedge angle distribution in the bottom panels
of Fig.~\ref{fig:experiment} can be discriminated from the optimum
more easily.  While the suboptimality of the cosine distribution is 
confirmed only for moderate to large variances $\Delta \hat{E} > 
0.75$, the wedge angular shape shows entirely different behavior: 
the uncertainty product increases with the variance.  This tendency 
of the wedge distribution can be readily explained: as the variance 
gets larger, more and more side maxima of the sinc-like angular 
momentum spectrum fall into the detected window $m\in [-15,15]$, 
yielding a corresponding grow of the uncertainty product.

\section{Conclusions}
\label{sec:conclusions}

In conclusion, we have formulated rigorous uncertainty relations for
angle and angular momentum based on circular variance as a proper 
statistical measure of angular error.  Fundamental Mathieu states 
were identified as intelligent states under the constraint of given 
uncertainty either in angle or in angular momentum.  In this sense, 
the Mathieu states provide the optimal distribution of information 
between the two observables with possible applications in information 
processing.  An optical test of the uncertainty relations was performed 
by using spatial light modulators both for the beam preparation and 
analysis.

Although the present experiment nicely confirmed our theory, the
resolution of the present setup was not sufficient for observing finer
details in the angular-momentum representation of light beams. Further
improvements both on the detection scheme and on hardware are highly
desirable. Our scheme is conceptually simple but suffers from
crosstalks and artifacts, especially at large helicities. New
detection schemes based on direct sensing of beam wavefront could
perhaps solve this problem. Concerning beam manipulation, spatial
light modulators used in our experiment, though very flexible and
easy-to-use devices, have also their drawbacks, namely small light
efficiencies and pixellated structures. A possible future upgrade of
the experimental setup lies in employing the optically-addressed SLM.

\acknowledgments

We acknowledge discussions with Andrei Klimov, Ioannes Rigas, and
Hubert de Guise.  This work was supported by the Czech Ministry of
Education, Projects MSM6198959213 and LC06007, the Czech Grant Agency,
Grant 202/06/307, and the Spanish Research Directorate, Grant
FIS2005-06714.

%\bibliography{angular}

\begin{thebibliography}{60}
\expandafter\ifx\csname natexlab\endcsname\relax\def\natexlab#1{#1}\fi
\expandafter\ifx\csname bibnamefont\endcsname\relax
  \def\bibnamefont#1{#1}\fi
\expandafter\ifx\csname bibfnamefont\endcsname\relax
  \def\bibfnamefont#1{#1}\fi
\expandafter\ifx\csname citenamefont\endcsname\relax
  \def\citenamefont#1{#1}\fi
\expandafter\ifx\csname url\endcsname\relax
  \def\url#1{\texttt{#1}}\fi
\expandafter\ifx\csname urlprefix\endcsname\relax\def\urlprefix{URL }\fi
\providecommand{\bibinfo}[2]{#2}
\providecommand{\eprint}[2][]{\url{#2}}

\bibitem[{\citenamefont{Pe{\v{r}}inova
  et~al.}(1998)\citenamefont{Pe{\v{r}}inova, Luk{\v{s}}, and
  Pe{\v{r}}ina}}]{Perinova:1998}
\bibinfo{author}{\bibfnamefont{V.}~\bibnamefont{Pe{\v{r}}inova}},
  \bibinfo{author}{\bibfnamefont{A.}~\bibnamefont{Luk{\v{s}}}},
  \bibnamefont{and}
  \bibinfo{author}{\bibfnamefont{J.}~\bibnamefont{Pe{\v{r}}ina}},
  \emph{\bibinfo{title}{Phase in Optics}} (\bibinfo{publisher}{World
  Scientific}, \bibinfo{address}{Singapore}, \bibinfo{year}{1998}).

\bibitem[{\citenamefont{Luis and S{\'a}nchez-Soto}(2000)}]{Luis:2000}
\bibinfo{author}{\bibfnamefont{A.}~\bibnamefont{Luis}} \bibnamefont{and}
  \bibinfo{author}{\bibfnamefont{L.~L.} \bibnamefont{S{\'a}nchez-Soto}},
  \bibinfo{journal}{Prog. Opt.} \textbf{\bibinfo{volume}{44}},
  \bibinfo{pages}{421} (\bibinfo{year}{2000}).

\bibitem[{\citenamefont{Kastrup}(2006)}]{Kastrup:2006}
\bibinfo{author}{\bibfnamefont{H.~A.} \bibnamefont{Kastrup}},
  \bibinfo{journal}{Phys. Rev. A} \textbf{\bibinfo{volume}{73}},
  \bibinfo{pages}{052104} (\bibinfo{year}{2006}).

\bibitem[{\citenamefont{Allen et~al.}(2003)\citenamefont{Allen, Barnett, and
  Padgett}}]{Allen:2003}
\bibinfo{author}{\bibfnamefont{L.}~\bibnamefont{Allen}},
  \bibinfo{author}{\bibfnamefont{S.~M.} \bibnamefont{Barnett}},
  \bibnamefont{and} \bibinfo{author}{\bibfnamefont{M.~J.}
  \bibnamefont{Padgett}}, \emph{\bibinfo{title}{Optical Angular Momentum}}
  (\bibinfo{publisher}{Institute of Physics Publishing},
  \bibinfo{address}{Bristol}, \bibinfo{year}{2003}).

\bibitem[{\citenamefont{Allen et~al.}(1992)\citenamefont{Allen, Beijersbergen,
  Spreeuw, and Woerdman}}]{Allen:1992}
\bibinfo{author}{\bibfnamefont{L.}~\bibnamefont{Allen}},
  \bibinfo{author}{\bibfnamefont{M.~W.} \bibnamefont{Beijersbergen}},
  \bibinfo{author}{\bibfnamefont{R.~J.~C.} \bibnamefont{Spreeuw}},
  \bibnamefont{and} \bibinfo{author}{\bibfnamefont{J.~P.}
  \bibnamefont{Woerdman}}, \bibinfo{journal}{Phys. Rev. A}
  \textbf{\bibinfo{volume}{45}}, \bibinfo{pages}{8185} (\bibinfo{year}{1992}).

\bibitem[{\citenamefont{Bouchal and \v{C}elechovsk\'{y}}(2004)}]{Bouchal:2004}
\bibinfo{author}{\bibfnamefont{Z.}~\bibnamefont{Bouchal}} \bibnamefont{and}
  \bibinfo{author}{\bibfnamefont{R.}~\bibnamefont{\v{C}elechovsk\'{y}}},
  \bibinfo{journal}{New J. Phys.} \textbf{\bibinfo{volume}{6}},
  \bibinfo{pages}{131} (\bibinfo{year}{2004}).

\bibitem[{\citenamefont{Gibson et~al.}(2004)\citenamefont{Gibson, Courtial,
  Padgett, Vasnetsov, Pas'ko, Barnett, and Franke-Arnold}}]{Gibson:2004}
\bibinfo{author}{\bibfnamefont{G.}~\bibnamefont{Gibson}},
  \bibinfo{author}{\bibfnamefont{J.}~\bibnamefont{Courtial}},
  \bibinfo{author}{\bibfnamefont{M.~J.}~\bibnamefont{Padgett}},
  \bibinfo{author}{\bibfnamefont{M.}~\bibnamefont{Vasnetsov}},
  \bibinfo{author}{\bibfnamefont{V.}~\bibnamefont{Pas'ko}},
  \bibinfo{author}{\bibfnamefont{S.~M.} \bibnamefont{Barnett}},
  \bibnamefont{and}
  \bibinfo{author}{\bibfnamefont{S.}~\bibnamefont{Franke-Arnold}},
  \bibinfo{journal}{Opt. Express} \textbf{\bibinfo{volume}{12}},
  \bibinfo{pages}{5448} (\bibinfo{year}{2004}).

\bibitem[{\citenamefont{{\v{C}}elechovsk{\'{y}} and
  Bouchal}(2007)}]{Celechovsky:2007}
\bibinfo{author}{\bibfnamefont{R.}~\bibnamefont{{\v{C}}elechovsk{\'{y}}}}
  \bibnamefont{and} \bibinfo{author}{\bibfnamefont{Z.}~\bibnamefont{Bouchal}},
  \bibinfo{journal}{New J. Phys.} \textbf{\bibinfo{volume}{9}},
  \bibinfo{pages}{328} (\bibinfo{year}{2007}).

\bibitem[{\citenamefont{Mair et~al.}(2001)\citenamefont{Mair, Vaziri, Weihs,
  and Zeilinger}}]{Mair:2001}
\bibinfo{author}{\bibfnamefont{A.}~\bibnamefont{Mair}},
  \bibinfo{author}{\bibfnamefont{A.}~\bibnamefont{Vaziri}},
  \bibinfo{author}{\bibfnamefont{G.}~\bibnamefont{Weihs}}, \bibnamefont{and}
  \bibinfo{author}{\bibfnamefont{A.}~\bibnamefont{Zeilinger}},
  \bibinfo{journal}{Nature (London)} \textbf{\bibinfo{volume}{412}},
  \bibinfo{pages}{313} (\bibinfo{year}{2001}).

\bibitem[{\citenamefont{Vaziri et~al.}(2002)\citenamefont{Vaziri, Weihs, and
  Zeilinger}}]{Vaziri:2002}
\bibinfo{author}{\bibfnamefont{A.}~\bibnamefont{Vaziri}},
  \bibinfo{author}{\bibfnamefont{G.}~\bibnamefont{Weihs}}, \bibnamefont{and}
  \bibinfo{author}{\bibfnamefont{A.}~\bibnamefont{Zeilinger}},
  \bibinfo{journal}{J. Opt. B} \textbf{\bibinfo{volume}{4}},
  \bibinfo{pages}{S47} (\bibinfo{year}{2002}).

\bibitem[{\citenamefont{Molina-Terriza
  et~al.}(2004)\citenamefont{Molina-Terriza, Vaziri, \v{R}eh\'a\v{c}ek, Hradil,
  and Zeilinger}}]{Molina:2004}
\bibinfo{author}{\bibfnamefont{G.}~\bibnamefont{Molina-Terriza}},
  \bibinfo{author}{\bibfnamefont{A.}~\bibnamefont{Vaziri}},
  \bibinfo{author}{\bibfnamefont{J.}~\bibnamefont{\v{R}eh\'a\v{c}ek}},
  \bibinfo{author}{\bibfnamefont{Z.}~\bibnamefont{Hradil}}, \bibnamefont{and}
  \bibinfo{author}{\bibfnamefont{A.}~\bibnamefont{Zeilinger}},
  \bibinfo{journal}{Phys. Rev. Lett.} \textbf{\bibinfo{volume}{92}},
  \bibinfo{pages}{167903} (\bibinfo{year}{2004}).

\bibitem[{\citenamefont{Langford et~al.}(2004)\citenamefont{Langford, Dalton,
  Harvey, O'Brien, Pryde, Gilchrist, Bartlett, and White}}]{Langford:2004}
\bibinfo{author}{\bibfnamefont{N.~K.} \bibnamefont{Langford}},
  \bibinfo{author}{\bibfnamefont{R.~B.} \bibnamefont{Dalton}},
  \bibinfo{author}{\bibfnamefont{M.~D.} \bibnamefont{Harvey}},
  \bibinfo{author}{\bibfnamefont{J.~L.} \bibnamefont{O'Brien}},
  \bibinfo{author}{\bibfnamefont{G.~J.} \bibnamefont{Pryde}},
  \bibinfo{author}{\bibfnamefont{A.}~\bibnamefont{Gilchrist}},
  \bibinfo{author}{\bibfnamefont{S.~D.} \bibnamefont{Bartlett}},
  \bibnamefont{and} \bibinfo{author}{\bibfnamefont{A.~G.} \bibnamefont{White}},
  \bibinfo{journal}{Phys. Rev. Lett.} \textbf{\bibinfo{volume}{93}},
  \bibinfo{pages}{053601} (\bibinfo{year}{2004}).

\bibitem[{\citenamefont{Oemrawsingh et~al.}(2005)\citenamefont{Oemrawsingh, Ma,
  Voigt, Aiello, Eliel, 't~Hooft, and Woerdman}}]{Oemrawsingh:2005}
\bibinfo{author}{\bibfnamefont{S.~S.~R.} \bibnamefont{Oemrawsingh}},
  \bibinfo{author}{\bibfnamefont{X.}~\bibnamefont{Ma}},
  \bibinfo{author}{\bibfnamefont{D.}~\bibnamefont{Voigt}},
  \bibinfo{author}{\bibfnamefont{A.}~\bibnamefont{Aiello}},
  \bibinfo{author}{\bibfnamefont{E.~R.} \bibnamefont{Eliel}},
  \bibinfo{author}{\bibfnamefont{G.~W.} \bibnamefont{'t~Hooft}},
  \bibnamefont{and} \bibinfo{author}{\bibfnamefont{J.~P.}
  \bibnamefont{Woerdman}}, \bibinfo{journal}{Phys. Rev. Lett.}
  \textbf{\bibinfo{volume}{95}}, \bibinfo{pages}{240501}
  (\bibinfo{year}{2005}).

\bibitem[{\citenamefont{Marrucci et~al.}(2006)\citenamefont{Marrucci, Manzo,
  and Paparo}}]{Marrucci:2006}
\bibinfo{author}{\bibfnamefont{L.}~\bibnamefont{Marrucci}},
  \bibinfo{author}{\bibfnamefont{C.}~\bibnamefont{Manzo}}, \bibnamefont{and}
  \bibinfo{author}{\bibfnamefont{D.}~\bibnamefont{Paparo}},
  \bibinfo{journal}{Phys. Rev. Lett.} \textbf{\bibinfo{volume}{96}},
  \bibinfo{pages}{163905} (\bibinfo{year}{2006}).

\bibitem[{\citenamefont{Molina-Terriza
  et~al.}(2007{\natexlab{a}})\citenamefont{Molina-Terriza, Torres, and
  Torner}}]{Molina:2007}
\bibinfo{author}{\bibfnamefont{G.}~\bibnamefont{Molina-Terriza}},
  \bibinfo{author}{\bibfnamefont{J.~P.} \bibnamefont{Torres}},
  \bibnamefont{and} \bibinfo{author}{\bibfnamefont{L.}~\bibnamefont{Torner}},
  \bibinfo{journal}{Nat. Phys.} \textbf{\bibinfo{volume}{3}},
  \bibinfo{pages}{305} (\bibinfo{year}{2007}{\natexlab{a}}).

\bibitem[{\citenamefont{Nieto}(1967)}]{Nieto:1967}
\bibinfo{author}{\bibfnamefont{M.~M.} \bibnamefont{Nieto}},
  \bibinfo{journal}{Phys. Rev. Lett.} \textbf{\bibinfo{volume}{18}},
  \bibinfo{pages}{182} (\bibinfo{year}{1967}).

\bibitem[{\citenamefont{Zak}(1969)}]{Zak:1969}
\bibinfo{author}{\bibfnamefont{J.}~\bibnamefont{Zak}}, \bibinfo{journal}{Phys.
  Rev.} \textbf{\bibinfo{volume}{187}} (\bibinfo{year}{1969}).

\bibitem[{\citenamefont{Whelan}(1980)}]{Whelan:1980}
\bibinfo{author}{\bibfnamefont{C.~T.} \bibnamefont{Whelan}},
  \bibinfo{journal}{J. Phys. A} \textbf{\bibinfo{volume}{13}},
  \bibinfo{pages}{L181} (\bibinfo{year}{1980}).

\bibitem[{\citenamefont{Loss and Mullen}(1992)}]{Loss:1992}
\bibinfo{author}{\bibfnamefont{D.}~\bibnamefont{Loss}} \bibnamefont{and}
  \bibinfo{author}{\bibfnamefont{K.}~\bibnamefont{Mullen}},
  \bibinfo{journal}{J. Phys. A} \textbf{\bibinfo{volume}{25}},
  \bibinfo{pages}{L235} (\bibinfo{year}{1992}).

\bibitem[{\citenamefont{Ohnuki and Kitakado}(1993)}]{Ohnuki:1993}
\bibinfo{author}{\bibfnamefont{Y.}~\bibnamefont{Ohnuki}} \bibnamefont{and}
  \bibinfo{author}{\bibfnamefont{S.}~\bibnamefont{Kitakado}},
  \bibinfo{journal}{J. Math. Phys.} \textbf{\bibinfo{volume}{34}},
  \bibinfo{pages}{2827} (\bibinfo{year}{1993}).

\bibitem[{\citenamefont{Szab{\'o} et~al.}(1994)\citenamefont{Szab{\'o}, Kis,
  Adam, and Janszky}}]{Szabo:1994}
\bibinfo{author}{\bibfnamefont{S.}~\bibnamefont{Szab{\'o}}},
  \bibinfo{author}{\bibfnamefont{Z.}~\bibnamefont{Kis}},
  \bibinfo{author}{\bibfnamefont{P.}~\bibnamefont{Adam}}, \bibnamefont{and}
  \bibinfo{author}{\bibfnamefont{J.}~\bibnamefont{Janszky}},
  \bibinfo{journal}{Quantum Opt.} \textbf{\bibinfo{volume}{6}},
  \bibinfo{pages}{527} (\bibinfo{year}{1994}).

\bibitem[{\citenamefont{Kosteleck{\'y} and Tudose}(1996)}]{Kostelecky:1996}
\bibinfo{author}{\bibfnamefont{V.~A.} \bibnamefont{Kosteleck{\'y}}}
  \bibnamefont{and} \bibinfo{author}{\bibfnamefont{B.}~\bibnamefont{Tudose}},
  \bibinfo{journal}{Phys. Rev. A} \textbf{\bibinfo{volume}{53}},
  \bibinfo{pages}{1978} (\bibinfo{year}{1996}).

\bibitem[{\citenamefont{Kowalski and Rembieli{\'n}ski}(2002)}]{Kowalski:2002}
\bibinfo{author}{\bibfnamefont{K.}~\bibnamefont{Kowalski}} \bibnamefont{and}
  \bibinfo{author}{\bibfnamefont{J.}~\bibnamefont{Rembieli{\'n}ski}},
  \bibinfo{journal}{J. Phys. A} \textbf{\bibinfo{volume}{35}},
  \bibinfo{pages}{1405} (\bibinfo{year}{2002}).

\bibitem[{\citenamefont{Bang and Berger}(2006)}]{Bang:2006}
\bibinfo{author}{\bibfnamefont{J.~Y.} \bibnamefont{Bang}} \bibnamefont{and}
  \bibinfo{author}{\bibfnamefont{M.~S.} \bibnamefont{Berger}},
  \bibinfo{journal}{Phys. Rev. D} \textbf{\bibinfo{volume}{74}},
  \bibinfo{pages}{125012} (\bibinfo{year}{2006}).

\bibitem[{\citenamefont{Franke-Arnold et~al.}(2004)\citenamefont{Franke-Arnold,
  Barnett, Yao, Leach, Courtial, and Padgett}}]{Franke-Arnold:2004}
\bibinfo{author}{\bibfnamefont{S.}~\bibnamefont{Franke-Arnold}},
  \bibinfo{author}{\bibfnamefont{S.~M.} \bibnamefont{Barnett}},
  \bibinfo{author}{\bibfnamefont{E.}~\bibnamefont{Yao}},
  \bibinfo{author}{\bibfnamefont{J.}~\bibnamefont{Leach}},
  \bibinfo{author}{\bibfnamefont{J.}~\bibnamefont{Courtial}}, \bibnamefont{and}
  \bibinfo{author}{\bibfnamefont{M.}~\bibnamefont{Padgett}},
  \bibinfo{journal}{New J. Phys.} \textbf{\bibinfo{volume}{6}},
  \bibinfo{pages}{103} (\bibinfo{year}{2004}).

\bibitem[{\citenamefont{Pegg et~al.}(2005)\citenamefont{Pegg, Barnett,
  Zambrini, Franke-Arnold, and Padgett}}]{Pegg:2005}
\bibinfo{author}{\bibfnamefont{D.~T.} \bibnamefont{Pegg}},
  \bibinfo{author}{\bibfnamefont{S.~M.} \bibnamefont{Barnett}},
  \bibinfo{author}{\bibfnamefont{R.}~\bibnamefont{Zambrini}},
  \bibinfo{author}{\bibfnamefont{S.}~\bibnamefont{Franke-Arnold}},
  \bibnamefont{and} \bibinfo{author}{\bibfnamefont{M.}~\bibnamefont{Padgett}},
  \bibinfo{journal}{New J. Phys.} \textbf{\bibinfo{volume}{7}},
  \bibinfo{pages}{62} (\bibinfo{year}{2005}).

\bibitem[{\citenamefont{Leach et~al.}(2002)\citenamefont{Leach, Padgett,
  Barnett, Franke-Arnold, and Courtial}}]{Leach:2002}
\bibinfo{author}{\bibfnamefont{J.}~\bibnamefont{Leach}},
  \bibinfo{author}{\bibfnamefont{M.~J.} \bibnamefont{Padgett}},
  \bibinfo{author}{\bibfnamefont{S.~M.} \bibnamefont{Barnett}},
  \bibinfo{author}{\bibfnamefont{S.}~\bibnamefont{Franke-Arnold}},
  \bibnamefont{and} \bibinfo{author}{\bibfnamefont{J.}~\bibnamefont{Courtial}},
  \bibinfo{journal}{Phys. Rev. Lett.} \textbf{\bibinfo{volume}{88}},
  \bibinfo{pages}{257901} (\bibinfo{year}{2002}).

\bibitem[{\citenamefont{Hradil et~al.}(2006)\citenamefont{Hradil, Rehacek,
  Bouchal, {\v{C}}elechovsk\'{y}, and S{\'{a}}nchez-Soto}}]{Hradil:2006}
\bibinfo{author}{\bibfnamefont{Z.}~\bibnamefont{Hradil}},
  \bibinfo{author}{\bibfnamefont{J.}~\bibnamefont{Rehacek}},
  \bibinfo{author}{\bibfnamefont{Z.}~\bibnamefont{Bouchal}},
  \bibinfo{author}{\bibfnamefont{R.}~\bibnamefont{{\v{C}}elechovsk\'{y}}},
  \bibnamefont{and} \bibinfo{author}{\bibfnamefont{L.~L.}
  \bibnamefont{S{\'{a}}nchez-Soto}}, \bibinfo{journal}{Phys. Rev. Lett.}
  \textbf{\bibinfo{volume}{97}} (\bibinfo{year}{2006}).

\bibitem[{\citenamefont{Guti{\'e}rrez-Vega
  et~al.}(2000)\citenamefont{Guti{\'e}rrez-Vega, Iturbe-Castillo, and
  Ch{\'a}vez-Cerda}}]{Gutierrez:2000}
\bibinfo{author}{\bibfnamefont{J.~C.} \bibnamefont{Guti{\'e}rrez-Vega}},
  \bibinfo{author}{\bibfnamefont{M.~D.} \bibnamefont{Iturbe-Castillo}},
  \bibnamefont{and}
  \bibinfo{author}{\bibfnamefont{S.}~\bibnamefont{Ch{\'a}vez-Cerda}},
  \bibinfo{journal}{Opt. Lett.} \textbf{\bibinfo{volume}{25}},
  \bibinfo{pages}{1493} (\bibinfo{year}{2000}).

\bibitem[{\citenamefont{Guti{\'e}rrez-Vega
  et~al.}(2003)\citenamefont{Guti{\'e}rrez-Vega, Rodr{\'{\i}}guez-Dagnino,
  Meneses-Nava, and Ch{\'a}vez-Cerda}}]{Gutierrez:2003}
\bibinfo{author}{\bibfnamefont{J.~C.} \bibnamefont{Guti{\'e}rrez-Vega}},
  \bibinfo{author}{\bibfnamefont{R.~M.}
  \bibnamefont{Rodr{\'{\i}}guez-Dagnino}},
  \bibinfo{author}{\bibfnamefont{M.~A.} \bibnamefont{Meneses-Nava}},
  \bibnamefont{and}
  \bibinfo{author}{\bibfnamefont{S.}~\bibnamefont{Ch{\'a}vez-Cerda}},
  \bibinfo{journal}{Am. J. Phys.} \textbf{\bibinfo{volume}{71}},
  \bibinfo{pages}{233} (\bibinfo{year}{2003}).

\bibitem[{\citenamefont{Bandr{\'e}s et~al.}(2004)\citenamefont{Bandr{\'e}s,
  Guti{\'e}rrez-Vega, and Ch{\'a}vez-Cerda}}]{Bandres:2004}
\bibinfo{author}{\bibfnamefont{M.~A.} \bibnamefont{Bandr{\'e}s}},
  \bibinfo{author}{\bibfnamefont{J.~C.} \bibnamefont{Guti{\'e}rrez-Vega}},
  \bibnamefont{and}
  \bibinfo{author}{\bibfnamefont{S.}~\bibnamefont{Ch{\'a}vez-Cerda}},
  \bibinfo{journal}{Opt. Lett.} \textbf{\bibinfo{volume}{29}},
  \bibinfo{pages}{44} (\bibinfo{year}{2004}).

\bibitem[{\citenamefont{Carruthers and Nieto}(1968)}]{Carruthers:1968}
\bibinfo{author}{\bibfnamefont{P.}~\bibnamefont{Carruthers}} \bibnamefont{and}
  \bibinfo{author}{\bibfnamefont{M.~M.} \bibnamefont{Nieto}},
  \bibinfo{journal}{Rev. Mod. Phys} \textbf{\bibinfo{volume}{40}},
  \bibinfo{pages}{411} (\bibinfo{year}{1968}).

\bibitem[{\citenamefont{Emch}(1972)}]{Emch:1972}
\bibinfo{author}{\bibfnamefont{G.~G.} \bibnamefont{Emch}},
  \emph{\bibinfo{title}{Algebraic Methods in Statistical Mechanics and Quantum
  Field Theory}} (\bibinfo{publisher}{Wiley}, \bibinfo{address}{New York},
  \bibinfo{year}{1972}).

\bibitem[{\citenamefont{L{\'e}vy-Leblond}(1976)}]{Levy:1976}
\bibinfo{author}{\bibfnamefont{J.~M.} \bibnamefont{L{\'e}vy-Leblond}},
  \bibinfo{journal}{Ann. Phys. (N.Y.)} \textbf{\bibinfo{volume}{101}},
  \bibinfo{pages}{319} (\bibinfo{year}{1976}).

\bibitem[{\citenamefont{Judge and Lewis}(1963)}]{Judge:1963}
\bibinfo{author}{\bibfnamefont{D.}~\bibnamefont{Judge}} \bibnamefont{and}
  \bibinfo{author}{\bibfnamefont{J.}~\bibnamefont{Lewis}},
  \bibinfo{journal}{Phys. Lett.} \textbf{\bibinfo{volume}{5}},
  \bibinfo{pages}{190} (\bibinfo{year}{1963}).

\bibitem[{\citenamefont{Louisell}(1963)}]{Louisell:1963}
\bibinfo{author}{\bibfnamefont{W.~H.} \bibnamefont{Louisell}},
  \bibinfo{journal}{Phys. Lett.} \textbf{\bibinfo{volume}{7}},
  \bibinfo{pages}{60} (\bibinfo{year}{1963}).

\bibitem[{\citenamefont{Mackey}(1963)}]{Mackey:1963}
\bibinfo{author}{\bibfnamefont{G.~W.} \bibnamefont{Mackey}},
  \emph{\bibinfo{title}{Mathematical Foundations of Quantum Mechanics}}
  (\bibinfo{publisher}{Benjamin}, \bibinfo{address}{New York},
  \bibinfo{year}{1963}).

\bibitem[{\citenamefont{Helstrom}(1976)}]{Helstrom:1976}
\bibinfo{author}{\bibfnamefont{C.}~\bibnamefont{Helstrom}},
  \emph{\bibinfo{title}{Quantum Detection and Estimation Theory}}
  (\bibinfo{publisher}{Academic Press}, \bibinfo{address}{New York},
  \bibinfo{year}{1976}).

\bibitem[{\citenamefont{Leonhardt et~al.}(1995)\citenamefont{Leonhardt,
  Vaccaro, B{\"o}hmer, and Paul}}]{Leonhardt:1995}
\bibinfo{author}{\bibfnamefont{U.}~\bibnamefont{Leonhardt}},
  \bibinfo{author}{\bibfnamefont{J.~A.} \bibnamefont{Vaccaro}},
  \bibinfo{author}{\bibfnamefont{B.}~\bibnamefont{B{\"o}hmer}},
  \bibnamefont{and} \bibinfo{author}{\bibfnamefont{H.}~\bibnamefont{Paul}},
  \bibinfo{journal}{Phys. Rev. A} \textbf{\bibinfo{volume}{51}},
  \bibinfo{pages}{84} (\bibinfo{year}{1995}).

\bibitem[{\citenamefont{Luis and S{\'a}nchez-Soto}(1998)}]{Luis:1998}
\bibinfo{author}{\bibfnamefont{A.}~\bibnamefont{Luis}} \bibnamefont{and}
  \bibinfo{author}{\bibfnamefont{L.~L.} \bibnamefont{S{\'a}nchez-Soto}},
  \bibinfo{journal}{Eur. Phys. J. D} \textbf{\bibinfo{volume}{3}},
  \bibinfo{pages}{195} (\bibinfo{year}{1998}).

\bibitem[{\citenamefont{Mumford}(1983)}]{Mumford:1983}
\bibinfo{author}{\bibfnamefont{D.}~\bibnamefont{Mumford}},
  \emph{\bibinfo{title}{Tata Lectures on Theta I}}
  (\bibinfo{publisher}{Birkhauser}, \bibinfo{address}{Boston},
  \bibinfo{year}{1983}).

\bibitem[{\citenamefont{Abramowitz and Stegun}(1984)}]{Abramowitz:1984}
\bibinfo{editor}{\bibfnamefont{M.}~\bibnamefont{Abramowitz}} \bibnamefont{and}
  \bibinfo{editor}{\bibfnamefont{I.~A.} \bibnamefont{Stegun}}, eds.,
  \emph{\bibinfo{title}{Handbook of Mathematical Functions}}
  (\bibinfo{publisher}{Dover}, \bibinfo{address}{New York},
  \bibinfo{year}{1984}).

\bibitem[{\citenamefont{Mardia and Jupp}(2000)}]{Mardia:2000}
\bibinfo{author}{\bibfnamefont{K.~V.} \bibnamefont{Mardia}} \bibnamefont{and}
  \bibinfo{author}{\bibfnamefont{P.~E.} \bibnamefont{Jupp}},
  \emph{\bibinfo{title}{Directional Statistics}} (\bibinfo{publisher}{Wiley},
  \bibinfo{address}{Chichester}, \bibinfo{year}{2000}).

\bibitem[{\citenamefont{Barakat}(1987)}]{Barakat:1987}
\bibinfo{author}{\bibfnamefont{R.}~\bibnamefont{Barakat}}, \bibinfo{journal}{J.
  Opt. Soc. Am. A} \textbf{\bibinfo{volume}{4}}, \bibinfo{pages}{1213}
  (\bibinfo{year}{1987}).

\bibitem[{\citenamefont{Klimov and Chumakov}(1997)}]{Klimov:1997}
\bibinfo{author}{\bibfnamefont{A.}~\bibnamefont{Klimov}} \bibnamefont{and}
  \bibinfo{author}{\bibfnamefont{S.}~\bibnamefont{Chumakov}},
  \bibinfo{journal}{Phys. Lett. A} \textbf{\bibinfo{volume}{235}},
  \bibinfo{pages}{7} (\bibinfo{year}{1997}).

\bibitem[{\citenamefont{Perelomov}(1986)}]{Perelomov:1986}
\bibinfo{author}{\bibfnamefont{A.}~\bibnamefont{Perelomov}},
  \emph{\bibinfo{title}{Generalized Coherent States and their Applications}}
  (\bibinfo{publisher}{Springer}, \bibinfo{address}{Berlin},
  \bibinfo{year}{1986}).

\bibitem[{\citenamefont{Gonz{\'a}lez and del Olmo}(1998)}]{Gonzalez:1998}
\bibinfo{author}{\bibfnamefont{J.~A.} \bibnamefont{Gonz{\'a}lez}}
  \bibnamefont{and} \bibinfo{author}{\bibfnamefont{M.~A.} \bibnamefont{del
  Olmo}}, \bibinfo{journal}{J. Phys. A} \textbf{\bibinfo{volume}{31}},
  \bibinfo{pages}{8841} (\bibinfo{year}{1998}).

\bibitem[{\citenamefont{Hall and Mitchell}(2002)}]{Hall:2002}
\bibinfo{author}{\bibfnamefont{B.~C.} \bibnamefont{Hall}} \bibnamefont{and}
  \bibinfo{author}{\bibfnamefont{J.~J.} \bibnamefont{Mitchell}},
  \bibinfo{journal}{J. Math. Phys.} \textbf{\bibinfo{volume}{43}},
  \bibinfo{pages}{1211} (\bibinfo{year}{2002}).

\bibitem[{\citenamefont{Kowalski et~al.}(1996)\citenamefont{Kowalski,
  Rembieli{\'n}ski, and Papaloucas}}]{Kowalski:1996}
\bibinfo{author}{\bibfnamefont{K.}~\bibnamefont{Kowalski}},
  \bibinfo{author}{\bibfnamefont{J.}~\bibnamefont{Rembieli{\'n}ski}},
  \bibnamefont{and} \bibinfo{author}{\bibfnamefont{L.~C.}
  \bibnamefont{Papaloucas}}, \bibinfo{journal}{J. Phys. A}
  \textbf{\bibinfo{volume}{29}}, \bibinfo{pages}{4149} (\bibinfo{year}{1996}).

\bibitem[{\citenamefont{Rao}(1965)}]{Rao:1965}
\bibinfo{author}{\bibfnamefont{C.~R.} \bibnamefont{Rao}},
  \emph{\bibinfo{title}{Linear Statistical Inference and its Applications}}
  (\bibinfo{publisher}{Wiley}, \bibinfo{address}{New York},
  \bibinfo{year}{1965}).

\bibitem[{\citenamefont{Breitenberger}(1985)}]{Breitenberger:1985}
\bibinfo{author}{\bibfnamefont{E.}~\bibnamefont{Breitenberger}},
  \bibinfo{journal}{Found. Phys.} \textbf{\bibinfo{volume}{15}},
  \bibinfo{pages}{353} (\bibinfo{year}{1985}).

\bibitem[{\citenamefont{Uffink}(1990)}]{Uffink:1990}
\bibinfo{author}{\bibfnamefont{J.~B.~M.} \bibnamefont{Uffink}}, Ph.D. thesis,
  \bibinfo{school}{University of Utrecht} (\bibinfo{year}{1990}).

\bibitem[{\citenamefont{Bialynicki-Birula
  et~al.}(1993)\citenamefont{Bialynicki-Birula, Freyberger, and
  Schleich}}]{Bialynicki:1993}
\bibinfo{author}{\bibfnamefont{I.}~\bibnamefont{Bialynicki-Birula}},
  \bibinfo{author}{\bibfnamefont{M.}~\bibnamefont{Freyberger}},
  \bibnamefont{and} \bibinfo{author}{\bibfnamefont{W.}~\bibnamefont{Schleich}},
  \bibinfo{journal}{Phys. Scr.} \textbf{\bibinfo{volume}{T48}},
  \bibinfo{pages}{113} (\bibinfo{year}{1993}).

\bibitem[{\citenamefont{Forbes and Alonso}(2001)}]{Forbes:2001}
\bibinfo{author}{\bibfnamefont{G.~W.} \bibnamefont{Forbes}} \bibnamefont{and}
  \bibinfo{author}{\bibfnamefont{M.~A.} \bibnamefont{Alonso}},
  \bibinfo{journal}{Am. J. Phys.} \textbf{\bibinfo{volume}{69}},
  \bibinfo{pages}{340} (\bibinfo{year}{2001}).

\bibitem[{\citenamefont{Bluhm et~al.}(1995)\citenamefont{Bluhm, Kosteleck{\'y},
  and Tudose}}]{Bluhm:1995}
\bibinfo{author}{\bibfnamefont{R.}~\bibnamefont{Bluhm}},
  \bibinfo{author}{\bibfnamefont{V.~A.} \bibnamefont{Kosteleck{\'y}}},
  \bibnamefont{and} \bibinfo{author}{\bibfnamefont{B.}~\bibnamefont{Tudose}},
  \bibinfo{journal}{Phys. Rev. A} \textbf{\bibinfo{volume}{52}},
  \bibinfo{pages}{2234 } (\bibinfo{year}{1995}).

\bibitem[{\citenamefont{McLachlan}(1947)}]{McLachlan:1947}
\bibinfo{author}{\bibfnamefont{N.~W.} \bibnamefont{McLachlan}},
  \emph{\bibinfo{title}{Theory and Application of {M}athieu Functions}}
  (\bibinfo{publisher}{Oxford University Press}, \bibinfo{address}{New York},
  \bibinfo{year}{1947}).

\bibitem[{\citenamefont{Opatrn{\'y}}(1995)}]{Opatrny:1995}
\bibinfo{author}{\bibfnamefont{T.}~\bibnamefont{Opatrn{\'y}}},
  \bibinfo{journal}{J. Phys. A} \textbf{\bibinfo{volume}{28}},
  \bibinfo{pages}{6961} (\bibinfo{year}{1995}).

\bibitem[{\citenamefont{Frenkel and Portugal}(2001)}]{Frenkel:2001}
\bibinfo{author}{\bibfnamefont{D.}~\bibnamefont{Frenkel}} \bibnamefont{and}
  \bibinfo{author}{\bibfnamefont{R.}~\bibnamefont{Portugal}},
  \bibinfo{journal}{J. Phys. A} \textbf{\bibinfo{volume}{34}},
  \bibinfo{pages}{3541} (\bibinfo{year}{2001}).

\bibitem[{\citenamefont{Vasnetsov et~al.}(2003)\citenamefont{Vasnetsov, Torres,
  Petrov, and Torner}}]{Vasnetsov:2003}
\bibinfo{author}{\bibfnamefont{M.~V.} \bibnamefont{Vasnetsov}},
  \bibinfo{author}{\bibfnamefont{J.~P.} \bibnamefont{Torres}},
  \bibinfo{author}{\bibfnamefont{D.~V.} \bibnamefont{Petrov}},
  \bibnamefont{and} \bibinfo{author}{\bibfnamefont{L.}~\bibnamefont{Torner}},
  \bibinfo{journal}{Opt. Lett.} \textbf{\bibinfo{volume}{28}},
  \bibinfo{pages}{2285} (\bibinfo{year}{2003}).

\bibitem[{\citenamefont{Molina-Terriza
  et~al.}(2007{\natexlab{b}})\citenamefont{Molina-Terriza, Rebane, Torres,
  Torner, and Carrasco}}]{Molina:2007b}
\bibinfo{author}{\bibfnamefont{G.}~\bibnamefont{Molina-Terriza}},
  \bibinfo{author}{\bibfnamefont{L.}~\bibnamefont{Rebane}},
  \bibinfo{author}{\bibfnamefont{J.~P.} \bibnamefont{Torres}},
  \bibinfo{author}{\bibfnamefont{L.}~\bibnamefont{Torner}}, \bibnamefont{and}
  \bibinfo{author}{\bibfnamefont{S.}~\bibnamefont{Carrasco}},
  \bibinfo{journal}{J. Eur. Opt. Soc.} \textbf{\bibinfo{volume}{2}},
  \bibinfo{pages}{07014} (\bibinfo{year}{2007}{\natexlab{b}}).

\end{thebibliography}

\end{document}